\documentclass[onecolumn,amsmath,showpacs,nofootinbib,12pt]{revtex4-1}
\usepackage{graphicx}
\usepackage{dcolumn}
\usepackage{bm}
\begin{document}
\newcommand{\hs}{\hspace*{0.5cm}}
\newcommand{\vs}{\vspace*{0.5cm}}
\newcommand{\be}{\begin{equation}}
\newcommand{\ee}{\end{equation}}
\newcommand{\bea}{\begin{eqnarray}}
\newcommand{\eea}{\end{eqnarray}}
\newcommand{\ben}{\begin{enumerate}}
\newcommand{\een}{\end{enumerate}}
\newcommand{\bde}{\begin{widetext}}
\newcommand{\ede}{\end{widetext}}
\newcommand{\nn}{\nonumber}
\newcommand{\crn}{\nonumber \\}
\newcommand{\Tr}{\mathrm{Tr}}
\newcommand{\non}{\nonumber}
\newcommand{\noi}{\noindent}
\newcommand{\al}{\alpha}
\newcommand{\la}{\lambda}
\newcommand{\bet}{\beta}
\newcommand{\ga}{\gamma}
\newcommand{\va}{\varphi}
\newcommand{\om}{\omega}
\newcommand{\pa}{\partial}
\newcommand{\+}{\dagger}
\newcommand{\fr}{\frac}
\newcommand{\bc}{\begin{center}}
\newcommand{\ec}{\end{center}}
\newcommand{\Ga}{\Gamma}
\newcommand{\de}{\delta}
\newcommand{\De}{\Delta}
\newcommand{\ep}{\epsilon}
\newcommand{\varep}{\varepsilon}
\newcommand{\ka}{\kappa}
\newcommand{\La}{\Lambda}
\newcommand{\si}{\sigma}
\newcommand{\Si}{\Sigma}
\newcommand{\ta}{\tau}
\newcommand{\up}{\upsilon}
\newcommand{\Up}{\Upsilon}
\newcommand{\ze}{\zeta}
\newcommand{\ps}{\psi}
\newcommand{\Ps}{\Psi}
\newcommand{\ph}{\phi}
\newcommand{\vph}{\varphi}
\newcommand{\Ph}{\Phi}
\newcommand{\Om}{\Omega}

\title{Neutrino masses and superheavy dark matter in the 3-3-1-1 model}
\author{D. T. Huong}
\email {dthuong@iop.vast.ac.vn} \affiliation{Institute of Physics, Vietnam Academy of Science and Technology, 10 Dao Tan, Ba Dinh, Hanoi, Vietnam}
\author{P. V. Dong}
\email {pvdong@iop.vast.ac.vn} \affiliation{Institute of Physics, Vietnam Academy of Science and Technology, 10 Dao Tan, Ba Dinh, Hanoi, Vietnam}
\date{\today}

\begin{abstract}

In this work, we interpret the 3-3-1-1 model when the $B-L$ and 3-3-1 breaking scales behave simultaneously as the inflation scale. This setup not only realizes the previously-achieved consequences of inflation and leptogenesis, but also provides new insights in superheavy dark matter and neutrino masses. We argue that the 3-3-1-1 model can incorporate a scalar sextet, which induces both small masses for the neutrinos via a combined type I and II seesaw and large masses for the new neutral fermions. Additionally, all the new particles have the large masses in the inflation scale. The lightest particle among the $W$-particles that have abnormal (i.e., wrong) $B-L$ number in comparison to those of the standard model particles may be a superheavy dark matter as it is stabilized by the $W$-parity. The dark matter candidate may be a Majorana fermion, a neutral scalar, or a neutral gauge boson, which was properly created in the early universe due to the gravitational effects on the vacuum or the thermal production after cosmic inflation.

\end{abstract}


\maketitle

\section{\label{intro}Introduction}
The $SU(3)_C\otimes SU(2)_L\otimes U(1)_Y$ standard model of strong and electroweak interactions with three quark and
 lepton families and a scalar doublet is an excellent description of the physics of our world down to $10^{-18}$ m order.
However, it also leaves many crucial questions of the nature unanswered~\cite{pdg}. Indeed, the standard model predicts only normal matter that occupies roundly 5\% mass-energy density of the universe. What remains beyond the standard model is about 25\% dark mater and 70\% dark energy. The standard model provides null masses for the neutrinos, but the experiments have proved that the
neutrinos have nonzero, small masses and flavor mixing. Besides, the standard model cannot solve the issues concerning
the early universe such as the baryon-number asymmetry and the inflationary expansion. On the theoretical side, the standard model cannot explain how the Higgs mass is stabilized against radiative corrections, why there are only three families of fermions, and what makes the electric charges be quantized.

Alternative to the popular proposals of grand unification, extradimension, and supersymmetry \cite{pdg}, a simple extension of the gauge symmetry to $SU(3)_C\otimes SU(3)_L\otimes U(1)_X\otimes U(1)_N$ (3-3-1-1) might address mumerous questions \cite{3311dm,3311ph,3311il,3311g,3311km}. Here, $SU(3)_L$ is an enlargement of the weak-isospin symmetry, while the last two factors determine the electric charge ($Q$) and baryon-minus-lepton number ($B-L$), respectively. The 3-3-1-1 model overhauls the mathematical and phenomenological aspects of the known 3-3-1 models \cite{331m1,331m2,331m3,331r1,331r2,331r3}. Indeed, $U(1)_N$ is necessarily included since $B-L$ does not commute and non-close algebraically with $SU(3)_L$. Consequently, $B-L$ and thus $N$ charge must be gauged, and the electroweak and $B-L$ interactions are unified similarly to the Glashow-Weinberg-Salam theory. The small neutrino masses can be achieved via seesaw mechanisms \cite{seesawi1,seesawi2,seesawi3,seesawi4,seesawi5,seesawii1,seesawii2,seesawii3} as a result of the 3-3-1-1 symmetry breaking. The dark matter candidates naturally appear as $W$-particles that possess abnormal (i.e., wrong) $B-L$ number, which transform nontrivially and are thus stabilized under the $W$-parity (like $R$-parity)---a remnant of the gauge symmetry unbroken by the vacuum. If the $U(1)_N$ breaking scale is large, the corresponding $U(1)_N$ breaking field could act as an inflaton, explaining the cosmological inflation. The $CP$-asymmetry decays of the right-handed neutrinos into normal matter or dark matter can generate the matter-antimatter asymmetry appropriately. The 3-3-1-1 model provides plausible solutions to the electric charge quantization and flavor problems. Particularly, the large flavor-changing neutral currents and potential $CPT$ violation due to the unwanted vacuums and interactions in the 3-3-1 model with right-handed neutrinos are excellently prevented.

In the 3-3-1-1 model \cite{3311dm}, the new neutral fermions $N_R$ have vanishing masses at the tree-level.
However, their masses can be generated by the effective operators that couple lepton triplets
$\psi_L$ to scalar triplet $\chi$. Such effective operators which are invariant under the gauge symmetry
and $W$-parity can be radiatively induced by the model itself.
Alternatively, the neutral fermion masses can be given at the tree-level by introducing their left-handed counterparts, $N_L$, which transform as gauge symmetry singlets, so-called the truly sterile particles \cite{3311ph}. In all cases discussed, the new particles of the corresponding 3-3-1 model including $N_R$ have masses in the 3-3-1 breaking scale. On the other hand, the observed neutrino masses in this model are
generated by a type I seesaw mechanism. It is naturally to impose the seesaw scale of $B-L$ breaking as the inflation scale, which is close to a hypothetical grand unification scale of $10^{16}$ GeV order \cite{gutscale1,gutscale2,gutscale3,gutscale4} (however, see Appdenix \ref{apda}), which is required for the successful inflation and leptogenesis scenarios \cite{3311il}. Hence, the remaining particles such as the inflaton, right-handed neutrinos, and $B-L$ gauge boson all pick up a mass in the inflation regime.

Let us ask which size the 3-3-1 breaking scale has? A possibility for it is at TeV scale as investigated in the literature \cite{3311il,331m1,331m2,331m3,331r1,331r2,331r3}. The new observation of this work is that it can be as large as the $B-L$ breaking scale associated with the seesaw and inflation ones.  Such large size for the 3-3-1 breaking scale is made available by the implement of a scalar sextet. This
new scalar sextet will couple to $\psi_L \psi_L$, which
consequently provides small masses for the neutrinos via a type II seesaw
mechanism, in addition to the type~I one. In contradiction to the previous proposals, the new neutral fermion masses are naturally large as given at the tree-level via the vacuum value of the scalar sextet, without necessarily acquiring either their sterile counterparts $N_L$ or the effective operators. The implication of the scalar sextet for lepton-flavor changing and leptogenesis processes is further hinted.
Despite of a previous study \cite{3311il}, the scalar sextet may decay
into two light leptons, possibly involving heavy lepton modes, which may
dominate over generated lepton number.  It is noteworthy that since the unitarity of the 3-3-1 model is cured as well as the proton stability is ensured \cite{3311g}, a large energy scale with regard to the 3-3-1 breaking is possible.

Interestingly enough, the dark matter candidates, which are the lightest particles among $W$-particles carrying abnormal $B-L$ numbers, are superheavy in the inflation regime, called superheavy dark matter \cite{pro1,pro2,pro3,pro4,pro5,pro6,pro7,pro8,pro9,pro10,pro11,pro12,pro13,th1,th2,th3,th4,th5,th6}. They are stabilized by the $W$-parity as a residual gauge symmetry. It is to be noted that the often-studied global symmetries could not keep the candidates stable since they are subsequently broken by the non-perturbative effects due to the gravitational anomalies \cite{global}. The superheavy dark matter candidates are suitable to be non-thermally generated, because by contrast the thermal relics should overclose the universe due to the unitarity constraint \cite{unitarity}. 

Let us recall that in the previous works \cite{3311dm,3311ph,3311il,3311g}, the $SU(3)_L$ symmetry breaking is at the TeV scale, which provides the dark matter candidates as thermal relics, limited below some hundreds of TeV.  Hence, the above proposal is an alternative solution to the dark matter question.  With the perspective of TeV dark matter, we hope that the search for thermal dark matter may be connected to
the discovery of new physics at TeV scale. In fact, there are the extensive experimental programs that set up to detect the thermal dark matter such as direct and indirect detections as well as accelerator searches.  However, none of these efforts have discovered a clear thermal dark matter and no evidence for new physics related to dark matter has been observed at the large hadron collider.  The lack of evidence of thermal dark matter candidates may provide an additional source of dark matter in form of non-thermal candidates.  The non-thermal dark matter candidates can provide the dominant source of dark matter and their self annihilation rates can be more larger than that of thermal dark matter. Therefore we do not only expect for experimental search but also other probes of the microscopic nature of dark matter \cite{SHDM1}. Specially if dark matter and scalar perturbations can grow during the non-thermal phase, an additional enhancement of dark matter sub-structure on the small scale and important implication for indirect detection signals as well as the process of structure formation are expected to obtain \cite{SHDM2}. 
On the other hand, if the existence of dark matter derives from the inflaton dynamics \cite{SHDM3}, it can be tested via measurements of inflationary parameter and/or the CMB isocurvature perturbations.

The rest of this work is organized as follows. In Sec. \ref{intr}, we
briefly review the 3-3-1-1 model,
introducing the scalar sextet and concentrating on its effects for the mass spectrum of neutrinos
and new fermions. Section \ref{sclsec} is devoted to the scalar potential
when including the contribution of the scalar sextet. We show that the type II seesaw scale appearing naturally small in the considering model. We also identify the dark matter candidates, gauge bosons, and their masses. The inflation and reheating are discussed in Sec \ref{them}. Section \ref{pheno} considers the lightest $W$-particle
as superheavy dark matter, and estimates their contribution to the present critical density, where the scenarios for superheavy dark mater production are briefly studied. Finally, we conclude this work and make outlooks in Sec \ref{concl}.

\section{\label{intr} The 3-3-3-1 model with scalar sextet}

Let $SU(2)_L$ extend to $SU(3)_L$. The $[SU(3)_L]^3$ anomaly does not vanish for each complex representation unlike $SU(2)_L$. The fundamental representations (triplets/antitriplets) of $SU(3)_L$ decompose as $3=2\oplus1$ and $3^*=2^*\oplus 1$ under $SU(2)_L$. Thus, all the left-handed fermion doublets will be embedded into $3$ or $3^*$, where for the second case $(f_2,-f_1)$ is an antidoublet, provided that $(f_1,f_2)$ is a doublet. Suppose that all the right-handed fermion singlets transform as $SU(3)_L$ singlets (note that they cannot be put in the above $3$ or $3^*$ except for leptons because $SU(3)_C$, $SU(3)_L$, and spacetime symmetry commute). Since the $[SU(3)_L]^3$ anomaly for $3$ and $3^*$ are opposite, this anomaly is cancelled out if the number of $3$ is equal that of $3^*$, which determines the number of families to match that of colors. Hence, the fermion representations under $SU(3)_L$ are arranged as given below, there $N_R$, $U$, $D$, and $\nu_R$ are new particles added to complete the representations as well as cancelling other anomalies. In principle, the new leptons $N_R$ may have arbitrary $Q$ and $B-L$ charges \cite{3311g}, but in this work we consider the simplest, nontrivial case, $Q(N_R)=[B-L](N_R)=0$ (their partners $N_L$ are thus gauge singlets, which are truly sterile and not imposed). The lepton triplets obey $Q=\mathrm{diag}(0,-1,0)$ and $B-L=\mathrm{diag}(-1,-1,0)$, which indicate that $Q$ and $B-L$ neither commute nor close algebraically with $SU(3)_L$. Hence, two new Abelian gauge groups arise as a result to close those symmetries by $SU(3)_C\otimes SU(3)_L\otimes U(1)_X\otimes U(1)_N$ (called 3-3-1-1), where the color group is also included for completeness, and $X,N$ respectively define $Q,B-L$ by the forms as obtained below when acting on a lepton triplet. The $Q$ and $B-L$ charges for new quarks are thus followed when acting such operators on quark triplets/antitriplets. Note that the left-handed and right-handed fermions have the same $Q$ and $B-L$. The $X$ and $N$ charges are determined as $X=\mathrm{Tr}(Q)/D$ and $N=\Tr(B-L)/D$, where $D$ is the dimension of corresponding $SU(3)_L$ representation.                         

The fermion content in the 3-3-1-1 model under consideration is given by
  \bea \psi_{aL} &=&
\left(\begin{array}{c}
               \nu_{aL}\\ e_{aL}\\ (N_{aR})^c
\end{array}\right) \sim (1,3, -1/3,-2/3),\label{fe1}\\
\nu_{aR}&\sim&(1,1,0,-1),\hs e_{aR} \sim (1,1, -1,-1),\label{fe2}
\\
Q_{\al L}&=&\left(\begin{array}{c}
  d_{\al L}\\  -u_{\al L}\\  D_{\al L}
\end{array}\right)\sim (3,3^*,0,0),\hs Q_{3L}=\left(\begin{array}{c} u_{3L}\\  d_{3L}\\ U_L \end{array}\right)\sim
 \left(3,3,1/3,2/3\right),\label{fe3}\\ u_{a
R}&\sim&\left(3,1,2/3,1/3\right),\hs d_{a R} \sim
\left(3,1,-1/3,1/3\right),\label{fe4}\\ U_{R}&\sim&
\left(3,1,2/3,4/3\right),\hs D_{\al R} \sim
\left(3,1,-1/3,-2/3\right),\label{fermion}\eea where $a=1,2,3$ and
$\al=1,2$ are family indices \cite{3311dm}. The quantum numbers in the
parentheses are provided upon the 3-3-1-1 subgroups, respectively.
The electric charge, baryon-minus-lepton charge, and
$W$-parity ($P$) are embedded in the 3-3-1-1 symmetry as \be
Q=T_3-\fr{1}{\sqrt{3}}T_8 + X,\hs B-L=-\fr{2}{\sqrt{3}}T_8+N,\hs
P=(-1)^{3(B-L)+2s}=(-1)^{-2\sqrt{3}T_8+3 N+2 s},\ee where $T_i$ $(i=1,2,3,...,8)$,
$X$, and $N$ are $SU(3)_L$, $U(1)_X$, and $U(1)_N$ charges,
respectively, and $s$ is spin. Additionally, we will denote the $SU(3)_C$ charges
as $t_i$. The new observation is that $B-L$ is a noncommuative gauge charge like $Q$, which is nontrivially unified with the weak forces, which is unlike the standard model $B-L$ symmetry. $W$-parity is nontrivial for the new particles that carry abnormal (wrong) $B-L$ charges unlike those defined for the standard model particles, called $W$-particles. The residual gauge operators $Q$ and $P$ are actually conserved by the vacuum. The new fermions $N_R$, $U$, and $D$ possess $(Q,B-L)$
as $(0,0)$, $(2/3,4/3)$, and $(-1/3,-2/3)$, respectively. Here, we see that they have $B-L$ unlike the ordinary leptons/quarks and are
$W$-odd, while all the ordinary fermions are $W$-even.

The fermion content as provided is also free from all the other anomalies. Indeed, the $[SU(3)_C]^3$ anomaly always vanishes since all the quarks are vector-like. Additionally, we have $X=Q-T_3+T_8/\sqrt{3}$ and $N=B-L+2 T_8/\sqrt{3}$, in which the anomalies as coupled to $Q$, $B-L$, and $T_{3,8}$ obviously vanish. Hence, the anomalies associated with $X,N$ are cancelled too. To see this explicitly, the nontrivial anomalies which make troublesome can be calculated as presented in Appendix \ref{ano3311}. Here, note that $\nu_R$ as supposed are in order to cancel the gravity anomaly $[\mathrm{gravity}]^2U(1)_N$ and the self-anomaly $[U(1)_N]^3$. Although the $B$ and $L$ charges are anomalous, regarding $B-L$ as a fundamental charge makes the model free from all the $B$ and $L$ anomalies. Further, it is easily to show that the anomalies are always cancelled, independent of the $Q$ and $B-L$ embedding coefficients in the gauge group, i.e. those charges of the new particles (cf. \cite{3311g}).  

The scalar content actually contains
 \bea \rho &=& \left(\begin{array}{c}
\rho^+_1\\
\rho^0_2\\
\rho^+_3\end{array}\right)\sim (1,3,2/3,1/3), \hs \hs  \eta =  \left(\begin{array}{c}
\eta^0_1\\
\eta^-_2\\
\eta^0_3\end{array}\right)\sim (1,3,-1/3,1/3),\label{vev1}\\
 \chi &=& \left(\begin{array}{c}
\chi^0_1\\
\chi^-_2\\
\chi^0_3\end{array}\right)\sim (1,3,-1/3,-2/3), \hs \hs  \phi \sim (1,1,0,2).\eea Here, the scalars $\eta_3$, $\rho_3$, and $\chi_{1,2}$ carry $B-L$ charge with one unit and are $W$-odd, whereas the remaining scalars possess $[B-L](\eta_{1,2}, \rho_{1,2}, \chi_3)=0$ and $[B-L](\phi)=2$ and are $W$-even. The vacuum expectation values (VEVs)
that conserve $Q$ and $P$ are obtained as \bea \langle \rho \rangle =
\fr{1}{\sqrt{2}}(0,v,0)^T,\hs \langle \eta \rangle =\fr{1}{\sqrt{2}}(u,0,0)^T,\hs \langle \chi
\rangle =\fr{1}{\sqrt{2}}(0,0,w)^T,\hs \langle \phi\rangle =\fr{1}{\sqrt{2}}\La.\label{vevss}
\eea The 3-3-1-1 symmetry is broken down to $SU(3)_C\otimes U(1)_Q\otimes U(1)_{B-L}$ due to $w,u,v$, while $U(1)_{B-L}$ is broken down to $P$ due to $\La$. Under the standard model symmetry we have three scalar doublets $(\rho_1,\rho_2)$, $(\eta_1,\eta_2)$, and $(\chi_1,\chi_2)$, where the third one is $W$-odd and integrated out. The first two are $W$-even, behaving in the weak scale, and the standard model like Higgs boson is a combination of $\rho_2$ and $\eta_1$.    

Observe that $N_{aR}$ are still massless at the renormalizable level. To generate the appropriate
masses for $N_{aR}$, we additionally introduce a scalar sextet, \bea S= \left(%
\begin{array}{ccc}
  S_{11}^0 & \fr{S_{12}^-}{\sqrt{2}}& \fr{S_{13}^0}{\sqrt{2}} \\
  \fr{S_{12}^-}{\sqrt{2}} & S_{22}^{--} & \fr{S_{23}^{-}}{\sqrt{2}}\\
  \fr{S_{13}^0}{\sqrt{2}}& \fr{S_{23}^{-}}{\sqrt{2}}& S^0_{33} \\
\end{array}%
\right)     \sim (1,6,-2/3,-4/3), \eea which couples to two $\psi_L$'s. The VEV of $S$ that conserves $W$-parity takes the form,  \bea
\left\langle S \right\rangle = \frac{1}{\sqrt{2}}\left(%
\begin{array}{ccc}
  \kappa& 0 & 0\\
  0 & 0& 0 \\
  0 & 0& \Delta \\
\end{array}%
\right).\eea Note that $S_{13}$ and $S_{23}$ have $B-L=-1$ and are $W$-odd, while the other components possess $[B-L](S_{11},S_{12},S_{22})=-2$, $[B-L](S_{33})=0$, and are $W$-even.

The Lagrangian of the considering model includes the ones in \cite{3311ph} (some parameters will be renamed for easily reading) plus the kinetic mixing term in \cite{3311km} and new contributions relevant to
the scalar sextet. Up to the gauge fixing and ghost terms, it is
given by
 \bea \mathcal{L}&=&\sum_{\mathrm{fermion\
multiplets}}\bar{\Psi}i\ga^\mu D_\mu \Psi + \sum_{\mathrm{scalar\ multiplets}}(D^\mu \Phi)^\dagger
(D_\mu \Phi)\crn && -\fr{1}{4}G_{i\mu\nu}G_i^{\mu\nu} -\fr 1 4 A_{i\mu\nu}A_i^{\mu\nu}-\fr 1 4
B_{\mu\nu} B^{\mu\nu}-\fr{1}{4}C_{\mu\nu} C^{\mu\nu}-\fr{\delta}{2} B_{\mu\nu}C^{\mu\nu}\crn &&-V(\rho,\eta,\chi,\phi,S) +
\mathcal{L}_{\mathrm{Yukawa}},\label{ttdd}\eea
where $D_\mu=\pa_\mu+ig_s G_{i\mu}t_i+ig A_{i\mu}T_i+ig_X B_\mu X+ig_N C_\mu N$ is covariant derivative. The field strength tensors, $G_{i\mu\nu}$, $A_{i\mu\nu}$, $B_{\mu\nu}$, and $C_{\mu\nu}$, are given as coupled to the gauge fields, $G_{i\mu}$, $A_{i\mu}$, $B_\mu$, and $C_\mu$, as well as the coupling constants, $g_s$, $g$, $g_X$, and $g_N$, of the 3-3-1-1 subgroups, respectively. The Yukawa Lagrangian is
\bea
\mathcal{L}_{\mathrm{Yukawa}}&=&h^e_{ab}\bar{\psi}_{aL}\rho e_{bR}
+h^\nu_{ab}\bar{\psi}_{aL}\eta\nu_{bR}+h'^\nu_{ab}\bar{\nu}^c_{aR}\nu_{bR}\phi +f_{ab} \bar{\psi}_{aL}^c S^\dagger \psi_{bL} \crn
&& +h^U\bar{Q}_{3L}\chi U_R + h^D_{\al \beta}\bar{Q}_{\al L} \chi^* D_{\beta R}+ h^u_a \bar{Q}_{3L}\eta u_{aR}+h^d_a\bar{Q}_{3L}\rho d_{aR} \crn
&&+ h^d_{\al a} \bar{Q}_{\al L}\eta^* d_{aR} +h^u_{\al a } \bar{Q}_{\al L}\rho^* u_{aR} + H.c.\label{ttddv}\eea The scalar potential is separated into two parts, $V(\rho,\eta,\chi, \phi,S)=V(\rho,\eta,\chi,\phi)+V(S)$, where \bea
V(\rho,\eta,\chi,\phi) &=& \mu^2_\phi \phi^\dagger
\phi  + \mu^2_\rho \rho^\dagger \rho + \mu^2_\chi \chi^\dagger \chi + \mu^2_\eta
\eta^\dagger \eta + \la (\phi^\dagger \phi)^2 + \la_1 (\rho^\dagger \rho)^2 \crn
&& + \la_2 (\chi^\dagger \chi)^2 + \la_3 (\eta^\dagger
\eta)^2+ \la_4 (\rho^\dagger \rho)(\chi^\dagger \chi) +\la_5 (\rho^\dagger
\rho)(\eta^\dagger \eta)\crn
&& +\la_6 (\chi^\dagger \chi)(\eta^\dagger \eta) +\la_7 (\rho^\dagger
\chi)(\chi^\dagger \rho) +\la_8 (\rho^\dagger \eta)(\eta^\dagger \rho)\crn
&&+\la_9 (\chi^\dagger
\eta)(\eta^\dagger \chi) + \la_{10} (\phi^\dagger
\phi)(\rho^\dagger\rho)+\la_{11}(\phi^\dagger \phi)(\chi^\dagger \chi)\crn
&&+\la_{12}(\phi^\dagger
\phi)(\eta^\dagger \eta) + (f_1 \epsilon^{mnp}\eta_m\rho_n\chi_p+H.c.), \label{scalar}\eea
\bea
V(S)&=& \mu_S^2 \Tr(SS^\dagger)+\zeta_{1}\Tr^2(SS^\dagger)+\zeta_{2} \Tr(SS^\dagger)^2 \crn
&& + (\zeta_{3} \eta^\dagger \eta
+\zeta_{4} \rho^\dagger \rho+\zeta_{5} \chi^\dagger \chi+\zeta_{6} \phi^\dagger \phi)\Tr(SS^\dagger)
\crn
&& + \zeta_{7} (\eta^\dagger S)(S^\dagger\eta)+\zeta_8 (\chi^\dagger S)(S^\dagger  \chi)+\zeta_9(\rho^\dagger S)(S^\dagger \rho)
\nonumber  \\ &&
+ (\zeta_{10} \eta^T S^\dagger \eta \phi^*+f_2 \chi^T S^\dagger \chi +H.c.). \label{scalar11}
\eea

To ensure that the scalar potential $V = V(\rho,\eta,\chi,\phi,S)$ is bounded from below (i.e., vacuum stability), the necessary conditions are 
\be \la>0,\hs \la_1>0,\hs \la_2>0,\hs \la_3>0,\hs \zeta_1+\zeta_2>0,\ee which could be obtained for $V>0$ when $\phi$, $\rho$, $\chi$, $\eta$, and $S$ separately tend to infinity, respectively. Additional conditions are $V>0$ for any two of the $\phi$, $\rho$, $\chi$, $\eta$, and $S$ fields simultaneously tending to infinity, which yield
\bea && \la_4+\la_7\theta(-\la_7)>-2\sqrt{\la_1\la_2}, \hs \la_5+\la_8\theta(-\la_8)>-2\sqrt{\la_1\la_3}, \crn 
&& \la_6+\la_9\theta(-\la_9)>-2\sqrt{\la_2\la_3},\hs \la_{10}>-2\sqrt{\la\la_1},\hs \la_{11}>-2\sqrt{\la\la_2},\crn 
&& \la_{12}>-2\sqrt{\la\la_3},\hs \zeta_6>-2\sqrt{\la(\zeta_1+\zeta_2)}, \hs \zeta_3+\zeta_7\theta(-\zeta_7)>-2\sqrt{\la_3(\zeta_1+\zeta_2)},\crn
&& \zeta_4+\zeta_9\theta(-\zeta_9)>-2\sqrt{\la_1(\zeta_1+\zeta_2)},\hs \zeta_5+\zeta_8\theta(-\zeta_8)>-2\sqrt{\la_2(\zeta_1+\zeta_2)},\eea where $\theta(x)$ is the Heaviside step function. Furthermore, $V>0$ for any three, any four, and the five of the $\phi$, $\rho$, $\chi$, $\eta$, and $S$ fields, respectively, simultaneously tending to infinity also provide extra conditions for vacuum stability. We might also have the constaints (but, most of them should be equivalent to the above conditions) for physical scalar masses as squared to be positive. On the other hand, to have desirable vacuum structure, i.e. the VEVs, the necessary conditions are $\mu^2_{\phi}<0$, $\mu^2_{\rho}<0$, $\mu^2_{\chi}<0$, $\mu^2_{\eta}<0$, and $\mu^2_{S}<0$.           

We see that the appearance of the scalar sextet does not affect the mass spectrum of the charged
leptons and quarks, which were presented in
\cite{3311dm}. Because the $W$-parity is conserved, i.e. $\langle S_{13}\rangle=\langle \eta_3\rangle =0$, the left-handed and right-handed neutrinos
do not mix with the neutral fermions, $N_{aR}$. The neutral fermions by
themselves couple to $S_{33}$ which yields their masses in $\Delta$
scale of the form, $-\frac{1}{2} \bar{N}_R m_N N^c_R + H.c.$,
where \bea [m_N]_{ab} =- \sqrt{2} f_{ab}\Delta, \eea which is different from the criteria in \cite{3311dm}. On the other hand, the left-handed neutrinos gain Majorana masses since they couple to $S_{11}$, $[m_L]_{ab}=-\sqrt{2}\kappa f_{ab}$. The right-handed neutrinos obtain Majorana masses because they interact with $\phi$, $[m_R]_{ab}=-\sqrt{2} \La h'^\nu_{ab}$. Whereas, the left-handed and right-handed neutrinos couple to $\eta_1$, so their Dirac masses are obtained as $[m^*_D]_{ab}=-uh^\nu_{ab}/\sqrt{2}$ \cite{3311dm}. Hence, the total mass Lagrangian for the neutrinos is  \bea
\mathcal{L}^\nu_{\mathrm{mass}}=-\frac{1}{2}\left(%
\begin{array}{cc}
  \bar{\nu}_L^c & \bar{\nu}_R \\
\end{array}%
\right)
m_\nu  \left(%
\begin{array}{c}
  \nu_L \\
  \nu_R^c \\
\end{array}%
\right)+H.c.,\eea
where $m_\nu$ has the form \bea
m_\nu =\left(%
\begin{array}{cc}
   m_L& m_D \\
  m^T_D & m_R \\
\end{array}%
\right). \eea

First note that $w, \Delta, \Lambda$ break $SU(3)_L\otimes U(1)_X\otimes U(1)_N$ down to $SU(2)_L\otimes U(1)_Y$ and provide the masses for the new particles, whereas $u,v, \kappa$ break $SU(2)_L\otimes U(1)_Y$ down to $U(1)_Q$ and give the masses for the standard model particles. To be consistent with the present data, we assume $u,v, \kappa \ll w,
\Lambda, \Delta $. In this limit, the scalar sextet shifts the $\rho$-parameter by $\Delta\rho\equiv \rho-1 \simeq - \frac{2 \kappa^2}{v^2+u^2} + \mathcal{O}[\kappa^4/(v^4,u^4),(u^2,v^2)/(w^2, \Lambda^2, \Delta^2)]$, which is negative. The positive contributions that come from the mass splittings of the fermion, scalar and vector doublets could make it overall positive and comparable to the global fit \cite{pdg}. We thus expect $2\kappa^2/(u^2+v^2)\sim 0.0004$, which implies $\kappa \sim 3.5\ \mathrm{GeV}$. Note that the $W$ mass can be approximated as $m^2_W\simeq
\frac{g^2}{4}(v^2+u^2)$, which yields $u^2+v^2
\simeq (246\ \mathrm{GeV})^2$, as used. Now that, due to the constraints, $\kappa\ll u,v \ll w,\La,\Delta$, thus $m_L\ll m_D\ll m_R$, the observed, light neutrinos $\sim \nu_L$ achieve masses via a combinational mechanism of type I and II seesaw, by \bea m_{\mathrm{light}}\simeq m_L-m_D m^{-1}_R m^T_D=
-\sqrt{2} \left[\kappa f - \frac{u^2}{4\Lambda} (h^{\nu})^* (h^{\prime
\nu})^{-1}(h^{\nu})^\dagger \right], \eea which are
naturally small since $\kappa$ and $u^2/\Lambda$ can be in eV scale, as shown below. The heavy neutrinos $\sim \nu_R$ have the masses, $m_{\mathrm{heavy}}\simeq
-\sqrt{2} h^{\prime \nu} \Lambda$, as retained, which are proportional to the $U(1)_N$ breaking scale, $\Lambda$.

We would like to emphasize that the VEVs
$u,v$ (including $\kappa$) break the electroweak symmetry. Whereas, the VEVs $w, \Delta$
break the $SU(3)_L\otimes U(1)_X\otimes U(1)_N$ symmetry, but they do not break the $U(1)_{B-L}$
symmetry, and well-known as the 3-3-1 scales. The VEV $\La$ (including $\kappa$) breaks $B-L$, thus $U(1)_N$ totally. It is naturally to suppose $w\sim \Delta$ and $u\sim v$, because they mainly break $SU(3)_L$ and $SU(2)_L$, respectively. The
phenomenological aspects of the 3-3-1-1 model can be divided into the corresponding regimes, such that
\ben
\item $w\sim \Lambda 
\sim \mathrm{TeV}$, as explicitly studied in \cite{3311ph}.
\item  $w \sim \mathrm{TeV}  \ll \Lambda \sim  m_{\mathrm{inflaton}}$,
as explicitly investigated in \cite{3311dm,3311il}.
\item $w\sim \La \sim m_{\mathrm{inflaton}}$, which is the new case under consideration.
\een
Below, we will show that both the 3-3-1 and $B-L$ breaking scales, $w$ and $\La$, can be kept at a very high energy scale as the inflation scale, which is close to a hypothetical grand unification scale (however, see Appendix \ref{apda} for extra discussions).
By this regime, it is best understood why the seesaw contributions, $\kappa$ and $u^2/\La$, are naturally small. The introduction of $S$, thus $\Delta$, provides (i) $N_{aR}$ are realized in the inflation energy regime, (ii) the neutrino masses of the type II seesaw fit the observed range in eV, and (iii) rich phenomenology in inflation, leptogenesis, and dark matter candidates. Of course, the leading conclusions of this work would remain if one omitted the scalar sextet.

\section {\label{sclsec}Scalar sector}

First of all, we recall that the considering model provides
the type II seesaw neutrino masses, given that $S_{11}$ has a tiny VEV, $\kappa$. Because the lepton number is a gauge charge, the Goldstone boson, well-known as Majoron, that is associated with this broken charge can be eliminated by the corresponding gauge field. There is no invisible decay mode of the $Z$ boson into the Majoron and its Higgs partner (however, see \cite{ma}). The Majoron problem is solved, which is unlike \cite{PB}. We will also show that $\kappa$ is naturally small, as suppressed and protected by the $B-L$ dynamics, due to the interaction, $-\zeta_{10} \eta^T S^\dagger \eta \phi^*$. Here, when $\phi$ gets a VEV, $\La$, it becomes $-\fr{1}{\sqrt{2}}\zeta_{10}\La \eta^T S^\dagger \eta $, which works as that in the theory of explicit lepton-number violation~\cite{ma}. The violation strength is set by $\La$.

Expanding the neutral scalars around their VEVs, we have
\bea \rho &=& \left(\begin{array}{c}
\rho^+_1\\
\frac{1}{\sqrt{2}}(v+S_2+iA_2)\\
\rho^+_3\end{array}\right),\hs
\eta = \left(\begin{array}{c}
\frac{1}{\sqrt{2}}(u+S_1+iA_1)\\
\eta^-_2\\
\frac{1}{\sqrt{2}}(S_3^\prime+iA_3^\prime)\end{array}\right),\crn
 \chi &=& \left(\begin{array}{c}
\frac{1}{\sqrt{2}}(S_1^\prime+i A_1^\prime)\\
\chi^-_2\\
\frac{1}{\sqrt{2}}(\om+S_3+iA_3)\end{array}\right),\hs
\phi = \frac{1}{\sqrt{2}}(\Lambda+S_4+i A_4),\eea and for the sextet,
\bea S= \left(%
\begin{array}{ccc}
  \frac{1}{\sqrt{2}}(\kappa+S_5+iA_5) & \frac{S^-_{12}}{\sqrt{2}}& \frac{1}{2}(S_2^\prime+iA_2^\prime) \\
  \frac{S^-_{12}}{\sqrt{2}} & S^{--}_{22} & \frac{S^-_{23}}{\sqrt{2}} \\
  \frac{1}{2}(S_2^\prime+iA_2^\prime) & \frac{S^-_{23}}{\sqrt{2}} & \frac{1}{\sqrt{2}}(\Delta+ S_6+iA_6) \\
\end{array}%
\right). \eea
Here, all the fields superscripted by a prime, $S^\prime$ and $A^\prime$, are $W$-odd, while the others, $S$ and $A$, are $W$-even. There is no mixing between two kinds of the fields, due to $W$-parity conservation. Also, the $W$-odd and $W$-even charged scalars do not mix.

The potential minimization
conditions are derived as \bea \sqrt{2}f_1 vw+u\left(\zeta_7
\kappa^2+2\zeta_{10} \kappa \Lambda+\lambda_{12}
\Lambda^2+\zeta_{3}(\kappa^2+\Delta^2)+2\lambda_3 u^2+\lambda_5
v^2+\lambda_6w^2+2\mu_\eta^2\right) &=&0, \label{T1}\crn
\sqrt{2}f_1 uw+v\left(2\lambda_1 v^2+\lambda_{10}
\Lambda^2+\zeta_{4}(\kappa^2+\Delta^2)+\lambda_4 w^2+ \lambda_5
u^2+ 2\mu_\rho^2\right)&=&0, \label{T2} \crn
\sqrt{2}f_1 uv+w\left(2\sqrt{2}f_2 \Delta+ \lambda_{11}\Lambda^2+
\zeta_{5} \kappa^2+\Delta^2(\zeta_8+\zeta_{5})+2\lambda_2
w^2+\lambda_4 v^2+\lambda_6 u^2+2\mu_\chi^2\right) &=&0, \label{T3} \crn
 \zeta_{10} u^2 \kappa+\Lambda\left(2\lambda
\Lambda^2+\lambda_{10}
v^2+\lambda_{11}w^2+\lambda_{12}u^2+\zeta_{6}(\kappa^2+\Delta^2)+2
\mu^2_\phi \right)&=& 0, \label{T4} \crn  u^2(\zeta_{10}
\Lambda+\kappa(\zeta_7+\zeta_{3}))+\kappa\left(2\zeta_{1}\Delta^2
+2(\zeta_{1}+\zeta_{2})\kappa^2+\zeta_{4}v^2+\zeta_{5}w^2+\zeta_{6}\Lambda^2
+2\mu_S^2
\right)&=&0, \label{T5} \crn
\sqrt{2}f_2w^2+\Delta\left(2\zeta_{1}\kappa^2+2(\zeta_{1}
+\zeta_{2})\Delta^2+\zeta_{3}u^2+\zeta_{4}v^2
+(\zeta_{5}+\zeta_8)w^2+\zeta_{6} \Lambda^2+2
\mu_S^2\right)&=&0. \nn \eea
To have the desirable vacuum, we set $\mu_\chi$, $\mu_\phi$, and $\mu_S$ in the inflation scale as mentioned. Correspondingly, the VEVs $w,\La,\Delta$ that reduce the 3-3-1-1 symmetry down to the standard model one are large in such regime. Indeed, for $\kappa,u,v=0$, we obtain
\bea && 2
\mu^2_\phi + \lambda_{11}w^2+ 2\lambda
\Lambda^2+\zeta_6\Delta^2 = 0, \crn
&& 2\mu^2_\chi+2\lambda_2
w^2+\lambda_{11}\Lambda^2+ (\zeta_8+\zeta_5) \Delta^2+ 2\sqrt{2}f_2 \Delta =0, \label{addh} \\
 &&2\mu_S^2+ (\zeta_8+\zeta_5)w^2+\zeta_6 \Lambda^2+2(\zeta_1
+\zeta_2)\Delta^2
+\sqrt{2}f_2w^2/\Delta = 0, \nn \eea which provide the $(w,\La,\Delta)$ solution proportionally to $(\mu_\chi,\mu_\phi,\mu_S)$, with an appropriate choice of the signs of the parameters. These three equations can also be deduced from the above six conditions if one uses $\mu_\eta,\mu_\rho,u,v,\kappa\ll \mu_\chi,\mu_\phi,\mu_S,w,\La, \Delta$.

At the low energy regime as of the standard model, all the heavy particles are integrated out. We come with the effective potential,
\bea V_{\mathrm{eff}}&=&\mu^2_\rho\rho^\dagger \rho+ \mu^2_\eta \eta^\dagger \eta +\la_1 (\rho^\dagger \rho)^2+\la_3(\eta^\dagger \eta)^2\crn
&&+\left(\la_5+f_1^2/\mu^2_\chi\right)(\rho^\dagger \rho)(\eta^\dagger \eta)+\left(\la_8-f_1^2/\mu^2_\chi\right)(\rho^\dagger \eta)(\eta^\dagger \rho), \eea where the last two terms received a contribution due to the $-f_1\eta\rho\chi$ interaction and its Hermitian conjugate; the other contributions are smaller and neglected. Note also that the fields $\eta$, $\rho$ denote only their doublet components, while their third components were integrated away. This potential yields the minimization conditions,\bea &&\mu^2_\eta+\la_3 u^2+\fr 1 2 \left(\la_5+f^2_1/\mu^2_\chi\right)v^2=0,\crn
&&\mu^2_\rho+\la_1 v^2+\fr 1 2 \left(\la_5+f^2_1/\mu^2_\chi\right)u^2=0, \eea that define the weak scales $(u,v)$, as usual. 

Also in this regime, since the left-handed neutrinos couple to the sextet by $f_{ab}\bar{\psi}^c_{aL}\psi_{bL} S^*$ and then the sextet couples to the standard model Higgs bosons by $-\zeta_{10}\eta\eta S^* \phi^*$, we obtain the effective interaction, \be -\fr{\La}{\sqrt{2}}\fr{1}{m^2_S}\zeta_{10}f_{ab} (\bar{\psi}^c_{aL}\eta^*)(\psi_{bL}\eta^*), \ee after integrating $S$ out as well as breaking the $B-L$ charge by $\langle \phi\rangle$ simultaneously. Here, $m_S$ denotes the mass of the scalar triplet located in the sextet, satisfying $m^2_S=\mu^2_S+\fr 1 2 \zeta_5 w^2 +\fr 1 2 \zeta_6 \La^2+\zeta_1 \Delta^2$. This interaction is responsible for the type II seesaw neutrino masses, \be [m_L]_{ab}=\fr{\La}{\sqrt{2}}\fr{u^2}{m^2_S}\zeta_{10}f_{ab},\ee which must agree with the result in the previous section. Indeed, the fifth minimization condition above implies a solution for $\kappa$,
\be \kappa=-\fr{\zeta_{10}u^2\La}{2m^2_S},\label{natural}\ee which matches the two results. It also implies $\kappa\sim u^2/\La$, which fits eV scale naturally.

In the pseudo-scalar sector, all the $W$-even fields, $A_1, A_2, A_3, A_4, A_5, A_6$, mix by themselves via the
mass matrix in such order as \bea \fr 1 2 M_A^2=\left(%
\begin{array}{cccccc}
  -\frac{\sqrt{2}f_1 vw+4\zeta_{10} u\kappa \Lambda}{4u} & -\frac{f_1 w}{2\sqrt{2}} & -\frac{f_1 v}{2\sqrt{2}}
   & \frac{\zeta_{10} u \kappa}{2} & \frac{\zeta_{10} u \Lambda}{2} & 0 \\
  -\frac{f_1 w}{2\sqrt{2}} & -\frac{f_1 uw}{2\sqrt{2}v} & -\frac{f_1 u}{2\sqrt{2}} & 0 & 0 & 0\\
  -\frac{f_1 v}{2\sqrt{2}} & -\frac{f_1 u}{2\sqrt{2}} & -\frac{f_1 uv+4f_2 \Delta w}{2\sqrt{2} w} &
  0 & 0 & \frac{f_2 w}{\sqrt{2}} \\
  \frac{\zeta_{10} u \kappa}{2} & 0 & 0 & -\frac{ \zeta_{10} u^2 \kappa}{4 \Lambda}
   & \frac{- \zeta_{10} u^2}{4} & 0 \\
  \frac{\zeta_{10} u \Lambda}{2} & 0 & 0 & -\frac{\zeta_{10}  u^2}{4}
  & -\frac{\zeta_{10}  u^2 \Lambda }{4\kappa} & 0 \\
  0 & 0 & \frac{f_2 w}{\sqrt{2}} & 0 & 0 & -\frac{f_2 w^2}{2\sqrt{2}\Delta} \\
\end{array}%
\right). \label{Ai} \eea
Also, in the scalar sector, all the $W$-even fields, $S_1, S_2,
S_3, S_4,S_5,S_6$, mix by themselves through a mass matrix, $\fr 1 2 M^2_S$, given in such order as
\bea
    \left(%
    \begin{array}{cccccc}
      -\frac{f_1 vw}{2\sqrt{2}u} +\la_3 u^2& \frac{\sqrt{2}}{4}(f_1 w+\sqrt{2}\la_5uv)
      &  \frac{\sqrt{2}}{4}(f_1 v+\sqrt{2}\la_6uw) & \frac{u}{2}(\zeta_{10} \kappa+\la_{12}
      \Lambda)
       & m^2_{{15}}& \frac{\zeta_{3}u \Delta}{2} \\
      \frac{\sqrt{2}}{4}(f_1 w+\sqrt{2}\la_5 uv) & -\frac{f_1 uw}{2\sqrt{2}v}+\la_1v^2 &
      \frac{\sqrt{2}}{4}(f_1 u+\sqrt{2}\la_4 vw) & \frac{\la_{10}v \Delta}{2} &
      \frac{\zeta_{4}v \kappa}{2}& \frac{\zeta_{4}v \Delta}{2} \\
     \frac{\sqrt{2}}{4}(f_1 v+\sqrt{2}\la_{6}uw) & \frac{\sqrt{2}}{4}(f_1 u+\sqrt{2}
     \la_4vw) &  -\frac{f_1 uv}{2\sqrt{2}w}+\la_2 w^2 &  \frac{\la_{11}w \Lambda}{2} & \frac{\zeta_{5}\kappa w}{2}
     & \fr{f_2 w}{\sqrt{2}}\\
     \frac{u}{2}(\zeta_{10} \kappa+\la_{12}\Lambda) &\frac{\la_{10}v\Lambda}{2}
      & \frac{\la_{11}w\Lambda}{2} & -\frac{\zeta_{10} \kappa u^2}{4 \Lambda}
      +\la \Lambda^2 & m^2_{{45}}
      &\frac{\zeta_{6}\Delta \Lambda}{2} \\
     m^2_{{15}}
      & \frac{\zeta_{4}\kappa v}{2}
       & \frac{\zeta_{5} \kappa w}{2}
       & m^2_{{45}}
        &m^2_{{55}}&\zeta_{1} \kappa \Delta \\
      \frac{\zeta_{3}u \Delta}{2} & \frac{\zeta_{4}v \Delta}{2} & \fr{f_2 w}{\sqrt{2}} &
       \frac{\zeta_{6}\Delta \Lambda}{2} & \zeta_{1}\kappa \Delta&m^2_{{66}} \\
    \end{array}%
\right),\nn
\label{neutral1}\eea
where we have defined,
\bea
   m^2_{{15}} &=&\frac{u}{2}(\zeta_{10}
                  \Lambda+\kappa(\zeta_7+\zeta_{3})), \hs
   m^2_{{55}}= -\frac{\zeta_{10} u^2 \Lambda}{4 \kappa}
  +(\zeta_{1}+\zeta_{2})\kappa^2 ,\\
 m^2_{{45}}&=&
   \frac{1}{4}(\zeta_{10} u^2+2 \zeta_{6}\kappa \Lambda),\hs
   m^2_{{66}}  = -\frac{f_2 w^2}{2\sqrt{2} \Delta} +(\zeta_{1}+\zeta_{2})\Delta^2.\eea

Above, we have investigated that the $S$, $\eta$, and $\phi$ interaction, i.e. $-\zeta_{10} \eta\eta S^*\phi^*$, is crucial to produce the observed neutrino masses as well as to make the model viable. Let us show this explicitly. First, we turn, by contrast, this interaction off, i.e. $\zeta_{10}=0$. The condition for the potential minimization in the $S_{11}$ direction becomes
\bea \kappa^2=-\frac{2 \mu^2_S+\zeta_{6}\La^2+\zeta_{5}w^2+2\zeta_{1}\Delta^2+(\zeta_7 +\zeta_{3})u^2+\zeta_{4}v^2}{2(\zeta_{1}+\zeta_{2})} \label{huong1}.\eea
Because $\mu_S,\Lambda,w, \Delta$ are proportional to the inflation scale, while $u,v$ are proportional to the weak scale, it is impossibly to impose a small value in eV for $\kappa$, unless unnatural fine-tunings among the two kinds of large scales are taken place. Thus, the $\kappa$ scale lies in the inflation energy regime, which ruins the standard model. Even if the fine-tuning is allowed, in this case, the pseudo-scalar mass matrix implies four massless fields. Three of them are the Goldstone bosons of the $Z,Z^\prime,C$ gauge bosons, such that
$G_Z\simeq \frac{1}{\sqrt{u^2+v^2}}(-u A_1+v
A_2), G_{Z^\prime} \simeq \frac{1}{\sqrt{w^2+ 4 \Delta^2}}(wA_3+2
\Delta A_6), G_{C}\simeq A_4$. The remaining massless field is $A_5$, which is a physical particle, acting similarly as a Majoron. On the other hand, the scalar mass matrix, $M^2_S$, also provides a physical partner of the Majoron, $S_5$, with mass $m^2_{S_5} \simeq (\zeta_{1}+\zeta_{2}) \kappa^2$, given at the leading order. Of course, this mass is as small as the neutrino mass. Therefore, the $Z$ boson would decay invisibly into $S_5 A_5$, having a rate equal to that of the $Z$ decay into two light neutrinos, which has experimentally been ruled out \cite{pdg}. By this view, the $\zeta_{10}$ coupling must be turned on, matching the fact that it conserves any symmetry of the theory and is renormalizable. Indeed, there is no reason why it is not presented in this model. 

Above, the presence of the $\zeta_{10}$ interaction may help us understanding why the type II seesaw neutrino masses are very tiny,
$\kappa \propto \frac{u^2}{\Lambda}$. This is because $\phi$ may play a role of inflaton field during the cosmological inflation
time, i.e. its VEV, $\Lambda$, is very large, in
$10^{13-14}$ GeV order \cite{3311il}, whereas $u,v$ are the electroweak scales, by which it obtains such a small mass $\kappa\sim \mathrm{eV}$. Note that the type I seesaw mechanism works analogously, where the mediators are right-handed neutrinos instead of the sextet, while $B-L$ is also broken by $\phi$ that directly couples to those right-handed neutrinos. See \cite{3311g} for details of the neutrino mass generation diagrams. Consequently, the natural small masses of the neutrinos might be originally correlated to the inflationary expansion of the early universe as all derived by the $\phi$ inflaton field. 
Furthermore, when $\zeta_{10} \neq 0$, the pseudo-scalar mass matrix (\ref{Ai}) yields that besides the three massless Goldstone bosons for $Z,Z^\prime,C$, the Majoron becomes massive. At the
leading order, the Majoron mass is given by \bea
 m^2_{A_5} \simeq -\frac{\zeta_{10} u^2 \Lambda}{2 \kappa}.
\eea The mass of its partner, $S_5$, is now \bea
 m^2_{S_5}\simeq -\frac{\zeta_{10} u^2 \Lambda}{2\kappa}.
\eea All these particles have mass in the inflation energy scale as expected.

Consider the $W$-odd scalars. The pseudo-scalar sector yields a massless state:
 \bea A_{1p}\simeq \frac{1}{\sqrt{w^2+2\Delta^2}}\left(w A_1^\prime +\sqrt{2}\Delta A_2^\prime \right),\eea
and two massive states with respective masses,
   \bea
       A_{2p}&\simeq & \frac{1}{\sqrt{w^2+2\Delta ^2}}\left( -\sqrt{2}\Delta
       A_{1}^\prime
       +w A_{2}^\prime \right),
       \hs m^2_{A_{2p}}\simeq -\frac{2\Delta^2+w^2}{4\Delta } \left(2\sqrt{2}f_2 +\zeta_{8}\Delta \right),\\
       A_{3p} &\simeq & A_3^\prime,\hs  m^2_{A_{3p}} \simeq  \frac{1}{2}\left[(\zeta_3+\zeta_7)\Delta^2-2\zeta_{10}\Delta \Lambda+\la_{12}\La^2+(\la_6+\la_9)w^2 \right].
 \eea Similarly, the scalar sector contains a massless state, called $S_{1p}$, and two massive states, named $S_{2p}$ and $S_{3p}$, determined as
   \bea
    S_{1p}&\simeq &\frac{1}{\sqrt{w^2+2\Delta^2}}\left(wS_1^\prime+\sqrt{2}\Delta S_2^\prime \right), \\
    S_{2p}&\simeq& \frac{1}{\sqrt{w^2+2\Delta^2}}\left(-\sqrt{2}\Delta S_1^\prime+w S_2^\prime \right), \\
    S_{3p}&\simeq & S_3^\prime,
    \eea with respective masses,
    \bea
    m_{S_{1p}}&=&0,\hs
    m^2_{S_{2p}} \simeq
   -\frac{2\Delta^2+w^2}{4\Delta} \left(2\sqrt{2}f_2 +  \zeta_{8}\Delta \right),\\
    m^2_{S_{3p}} &\simeq& \frac{1}{2}\left[(\zeta_3+\zeta_7)\Delta^2+2\zeta_{10} \Delta \Lambda+\la_{12}\La^2+(\la_6+\la_9) w^2 \right].
    \eea
Observe that the fields, $S_{1p}$ and $A_{1p}$, are the Goldstone bosons of the real and imaginary parts of the neutral, non-Hermitian $X$ gauge boson, respectively. Hence their combination, $G_X=\fr{1}{\sqrt{2}}(S_{1p}+iA_{1p})$, forms the Goldstone boson of $X$. Furthermore, $S_{2p}$ and $A_{2p}$ have
the same mass. They can be identified as a physical neutral complex field, $H^\prime = \frac{1}{\sqrt{2}} \left( S_{2p}+iA_{2p}\right)$, which is orthogonal to $G_X$.

Consider the $W$-odd, charged scalars. There are
 two massless Goldstone bosons, $G^\pm_{Y}$, as associated with the $Y^{\pm}$ gauge
 bosons, and four massive charged Higgs bosons, $H^\pm_{p1,p2}$. In the limit
 $\Lambda, \Delta, w \gg u,v,\kappa$, their eigenstates and masses
can be approximated as
 \bea
 G_{Y}^\pm &=& \frac{1}{\sqrt{2\Delta^2+w^2}}\left(
  w \chi_2^\pm
 +\sqrt{2}\Delta S^{\pm}_{23} \right), \hs m_{G_{Y}}=0,\\
 H^\pm_{p1} &= & \frac{1}{\sqrt{2\Delta^2+w^2}}\left(
  -\sqrt{2} \Delta \chi_2^\pm
 +w S^{\pm}_{23} \right),\hs  m_{H_{p1}}^2=
 -\frac{(2\Delta^2+w^2)(2\sqrt{2}f_2+\zeta_8 \Delta)}{4\Delta}, \\
 H^\pm_{p2} &= & \rho_3^\pm, \hs m_{H_{p2}}^2 =\frac{1}{2}\left[(\la_4+\la_7) w^2+(\zeta_4+\zeta_9) \Delta^2
+\zeta_{10}\La^2\right].
 \eea The doubly-charged scalars, $S^{\pm\pm}_{22}$, are physical fields by themselves and have large masses,
  \bea
      m^2_{S_{22}}=-\frac{\sqrt{2}f_2w^2+2\zeta_2 \Delta^3+\zeta_8 \Delta w^2}{2\Delta}.
\eea

Let us note that the sextet affects negligibly to the mass spectrum of the non-Hermitian $W$, $X$, $Y$ gauge bosons, as identified in \cite{3311ph}. Their states are \be W^\pm=\fr{1}{\sqrt{2}}(A_1\mp i A_2),\hs X^{0,0*}=\fr{1}{\sqrt{2}}(A_4\mp i A_5),\hs Y^\mp=\fr{1}{\sqrt{2}}(A_6\mp i A_7),\ee with respective masses,
\be m^2_W\simeq \fr{g^2}{4}(u^2+v^2),\hs m_X\simeq m_Y\simeq \fr{gw}{2}. \ee
The neutral gauge bosons, $A_{3},\ A_{8},\ B,\ C$, mix, as given in \cite{3311km}. And, this mass spectrum would be changed due to the contribution of the sextet. However, because of the limit, $u,v,\kappa \ll w, \Delta, \Lambda$, the $Z$ boson decouples (i.e. mixes infinitesimally) from the heavy $Z^\prime, C$ bosons, with mass $m^2_Z\simeq m^2_W/c^2_W$, whereas the $Z^\prime, C$ bosons may largely mix due to the contributions of $w,\La,\Delta$ and the kinetic mixing parameter.
Lastly, the $\rho$-parameter can be derived due to the contribution of $\kappa$, as mentioned before.

Further, from the potential minimization in $S_{11}$ and $S_{33}$ directions, we obtain a condition,
\bea
     \frac{f_2 w^2}{\sqrt{2} \Delta} -\frac{\zeta_7 u^2}{2}
     +\frac{\zeta_8 w^2}{2} -\frac{\zeta_{10} \Lambda u^2}{2 \kappa}
     +\zeta_{2}(\Delta^2-\kappa^2)=0.  \label{T56} \eea
Combining (\ref{T56}) and (\ref{natural}), we find that at least one of the two 3-3-1 breaking scales, $w$ or $\Delta$, must have the same magnitude as the $B-L$ breaking scale, $\Lambda$. It may also be derived from the three equations for the large scales in (\ref{addh}). Therefore, relaxing the condition, $w\sim\Delta$,  as above proposed, it leads to three hypotheses as follows
\ben
\item $w \sim O(1)\ \mathrm{TeV} \ll \Delta \sim \Lambda$. In this case, the mass spectrum
of the new particles is separated into two parts: The new gauge bosons $X^0, Y^\pm$, the exotic quarks, and some new Higgs (but not $H'$, $A'_3$, and $S'_3$) live in the TeV scale. Whereas, some other new Higgs including $H'$, $A'_3$, and $S'_3$, neutral fermions $N_R$, and new gauge bosons, $Z^\prime, C$, are heavy, with masses close to the inflation scale. This scenario does not provide any dark matter candidate, since the $X^0$ abundance completely vanishes, i.e. it annihilates totally before freeze-out \cite{3311dm}. See also \cite{split1,split2,split3} for other proposals.
\item $\Delta \sim O(1)\ \mathrm{TeV} \ll w \sim \Lambda$. All the new gauge bosons, exotic quarks, and most new Higgs bosons, including the $W$-odd scalars, gain masses in the inflation scale. Exclusively, the neutral fermions $N_R$ have mass in TeV scale. The lightest $N_R$ can be a thermal dark matter candidate \cite{3311dm}.
\item  $w\sim \Delta\sim \La$ being in the inflation scale.
The considering model induces non-thermal superheavy dark
matter (see, for other proposals, \cite{th1,th2,th3,th4,th5,th6}), because there are two necessary conditions: (i) the candidate is the lightest $W$-odd particle, LWP, which is stabilized by $W$-parity, (ii) the candidate was not in thermal equilibrium
with the cosmic plasma, since by contrast, it could overclose the universe due to the unitarity condition \cite{unitarity}. Hence, such candidate would be produced by various mechanisms for non-thermal relics \cite{pro1,pro2,pro3,pro4,pro5,pro6,pro7,pro8,pro9,pro10,pro11,pro12,pro13}. We will show that if its mass is as large as the inflation scale, it can be created by the gravitational mechanism, which is common in most models. If it has a smaller mass, just above the reheating temperature, it is naturally produced by the inflaton decay or thermal fusion. This observation is an interesting alternative connecting the 3-3-1-1 model to the physics at the early stage of the universe. Depending on the parameter space, the superheavy dark matter or LWP may be a neutral fermion (a combination of $N_{aR}$), a scalar (among $H'$, $S'_3$, and $A'_3$), or possibly a $X^0$ gauge boson. It is noteworthy that a non-thermal relic for the last one is viable, which is unlike its previous variant \cite{3311dm}.
\een

Before examining the superheavy dark matter, it is necessary to obtain the consistent inflation scenarios in order to fix the inflation scale, inflaton mass, and reheating temperature. Let us stress again that the inflation presenting in the current model is substantially different from the previous study \cite{3311il}.   

\section{ \label{them}Inflation and reheating}
 
We would like to note that the scalar fields, singlet $\phi$, triplet $\chi$, and sextet $S$, can have large VEVs, $\Lambda, w, \Delta$, respectively, proportional to the inflation scale of the early universe. At this energy scale, all the mentioned scalar fields can play the role as inflaton field(s) deriving the cosmic inflation (see, for an example, \cite{3311il}). Indeed, we can have a single-field inflation scenario as governed by one combination of the scalars or multi-field inflation scenarios as cooperated by a number of the combinations of the scalars, in the field space. Recall that $\chi_3$ and $S_{33}$ break $SU(3)_L\otimes U(1)_X\otimes U(1)_N$ down to $SU(2)_L\otimes U(1)_Y\otimes U(1)_{B-L}$, whereas $\phi$ breaks both $U(1)_N$ and $U(1)_{B-L}$ down to $W$-parity since $N(\phi)=[B-L](\phi) = 2\neq 0$, where note that these breakings as indicated is effectively translated to $S_{11}$ which along with $\eta_1$ and $\rho_2$ break both the electroweak symmetry and $B-L$. That said, the inflation may be related to the first kind symmetry breaking (i.e., 3-3-1 breaking) due to $\chi_3$, $S_{33}$ and/or the second kind symmetry breaking (i.e., $B-L$ breaking) due to $\phi$.   

Let us first consider a single-field inflation scenario linked to the $U(1)_{B-L}$
symmetry breaking as driven by the singlet $\phi$. The inflaton sector which is impacted from the model's potential in (\ref{scalar}) and (\ref{scalar11}) is thus extracted as
\bea
V_{\mathrm{tot}} &=& \mu^2_\phi \phi^\dag \phi +\la (\phi^\dag \phi)^2+\la_{11}(\phi^\dag \phi)(\chi^\dag \chi)+\zeta_6(\phi^\dagger\phi)\Tr(S^\dagger S)\crn
&& + \mu^2_\chi \chi^\dagger \chi +\la_2 (\chi^\dagger \chi)^2 + \mu^2_S\Tr (S^\dagger S) +\zeta_1[\Tr(S^\dagger S)]^2+\zeta_2 \Tr[(S^\dagger S)^2] \crn &&+ \zeta_5 (\chi^\dag \chi) \mathrm{Tr}(S S^\dag) +\zeta_8(\chi^\dagger S)(S^\dagger \chi)+ (f_2 \chi^T S^\dag \chi +H.c.),
\label{infla1}\eea
where $\phi$ is the inflaton field involving during inflation, while $\chi, S$ may be the water-fall fields. One might also include $\eta,\rho$ as water-fall fields, but they are radically light, subdominant, and thus omitted. During inflation, the inflaton potential reads
$V_{\mathrm{inflation}} = \mu^2_\phi \phi^\dag \phi +\la (\phi^\dag \phi)^2$, while the interactions of $\phi$ with $\chi, S$ as well as the self-terms of $\chi, S$ might terminate the inflation, where the inflation ends due to an instability triggered by $\phi$ when it reaches a critical value determined by the largest scalar mass between $\chi$ and $S$ \cite{alinde11}. As associated with the $B-L$ breaking, the inflaton slowly rolls down to the potential minimum from above $\phi>\La/\sqrt{2}$, and the inflation ends corresponding to a 3-3-1 symmetry breaking\footnote{By contrast, when the inflaton rolls down to the potential minimum from below $\phi<\La/\sqrt{2}$, the duration of inflation until end recognizes a 3-3-1 symmetry restoration. A dedicated study might be worth, but it is out of the scope of this work.}.      

As specified in \cite{3311il}, all the scalar couplings would be constrained to be radically small under the present data. Therefore, we further consider the inflaton potential to be radiatively induced as an effective 
potential, due to the interactions of $\phi$ with the $U(1)_N$ gauge boson ($C$), the scalar fields ($\phi,\chi,S$), and the right-handed neutrinos ($\nu_R$). To be concrete, we denote the inflaton as $\Phi=\sqrt{2}\Re(\phi)$, while $\Im(\phi)$ is a Goldstone boson which could be gauged away. We parametrize the effective potential in the leading-log approximation as \cite{effpot}
\be V(\Phi)\simeq  \fr {\la}{4}\left(\Phi^2-\La^2\right)^2+ \fr{a}{64\pi^2}\Phi^4 \ln \fr {\Phi^2}{\La^2} +V_0,  \ee
where \be a= -8 \sum_{i=1}^3 (h^{ \prime \nu}_{ii})^4 + 48 g_N^4+\fr 1 4 (\la_{11}^2+\zeta_6^2)+36\la^2\simeq 8[6g^4_N-\sum_{i=1}^3 (h^{ \prime \nu}_{ii})^4],\ee 
with $h'^\nu$ assumed to be flavor-diagonal, and the renormalization scale is fixed at $\La^2=-\mu^2_\phi/\la$, which is compatible to $w^2$ and $\Delta^2$, but should be significantly larger than $\mu^2_{\chi,S}$. The effective potential reveals a consistent local minimum if $a/\la>-63.165$. Provided that $\La$ is bounded below the Planck scale, the effective potential is governed by the quartic and log terms. Otherwise, when one put $\La$ beyond the Planck scale, the Coleman-Weinberg corrections would be negligible, and the tree-level potential dominates.  

We note that the inflation occurs as the inflaton slowly rolls down to the potential minimum at $\Phi \sim \La$. The slow roll parameters read \bea \epsilon(\Phi)=\frac{1}{2}m_P^2 \left( \frac{V^\prime}{V}\right)^2, \hs
\eta(\Phi)=m_P^2 \left( \frac{V^{\prime \prime}}{V}\right), \hs \xi^2 (\Phi)=m_P^4 \frac{V^\prime
V^{\prime \prime \prime}}{V^2}, \eea which satisfy $\epsilon (\Phi) \ll 1,\ \eta(\Phi)
\ll 1,\ \xi(\Phi) \ll 1$, where $m_P=(8\pi G_N)^{-1/2}\simeq 2.4 \times 10^{18}$ GeV is the reduced Planck mass. The spectral index $n_s$, the tensor-to-scalar ratio $r$, and the running index $\al $ can be approximated as
\bea n_s\simeq 1-6\epsilon+2\eta, \hs r\simeq 16\epsilon, \hs \al \simeq 16\epsilon\eta-24\epsilon^2-2\xi^2. \eea The experimental bounds for these quantities were summarized in \cite{pdg} as $n_s=0.968\pm0.006$, $\al=-0.003\pm0.007$, and $r<0.07$. The curvature perturbation is determined by \bea \triangle^2_\mathcal{R}=
\frac{V}{24\pi^2m_P^4\epsilon(\Phi)}, \eea which satisfies  $\triangle^2_\mathcal{R}=2.215 \times 10^{-9}$ at the pivot scale
$k_0=0.05\ \mathrm{Mpc}^{-1}$ to fit the CMB measurements \cite{pdg}. The  number of
e-folds is \bea N=\fr{1}{\sqrt{2}m_P} \int^{\Phi_0}_{\Phi_e}\frac{d \Phi}{\sqrt{\epsilon(\Phi)}}, \eea where $\Phi_e$ is given at the end of inflation specified by $\epsilon(\Phi_e)\simeq 1$, and $\Phi_0$ is given at the horizon exit corresponding to $k_0$, and $N=60$ is taken regarding a large inflation scale. 

The $\la$ coupling is determined from the $\triangle^2_{\mathcal{R}}$ constraint, which yields $\la\sim 10^{-12}$--$10^{-11}$ in the actual parameter regime. We are left with $r,n_s,\al$ correlatively related as functions of $\Phi$ involving from $\Phi_0$ to $\Phi_e$ for fixed values of the parameters, $a'=a/\la>-63.165$ and $\La$ selected in the range $(10^{13}\ \mathrm{GeV},m_P)$. As numerically evaluated, the values of $n_s$ and $r$ in agreement with the experimental constraints make the effective coupling reasonably large, $-60<a^\prime < -20$, and the $B-L$ breaking scale typically recovered in $\La\sim 10^{14}$--$10^{18}$ GeV. Since $a'$ is a function of the various couplings, the present inflation scenario does not constrain solely the gauge coupling $g_N$. However, the strength is set as $g_N\sim h'^\nu\sim (\la|a'|)^{1/4}\sim 10^{-2.5}$, which is quite smaller than the electroweak couplings. Therefore, it is seemingly impossible to choose the values of the Yukawa couplings so that $g_N$ is compatible with a unified gauge coupling by small extent responsible for a hypothetical, higher gauge symmetry if one proposes to search for (further, see Appendix \ref{apda}). 

The effective potential provides a VEV, $\langle \Phi \rangle \sim \La$, from the minimization condition $V'=0$. The inflaton mass is given at this VEV as \be m_\Phi=\sqrt{V''}=\sqrt{2\la\La^2+\fr{a}{8\pi^2}\langle \Phi\rangle^2}\sim \sqrt{\la}\La\sim 10^{8}-10^{12}\ \mathrm{GeV},\ee which should be smaller than the $U(1)_N$ gauge boson mass, $m_C=2g_N\langle \Phi\rangle$, due to $\sqrt{\la}\ll g_N$. If one supposes hierarchical Yukawa couplings, $h'^\nu_{11}\ll h'^\nu_{22,33}\sim g_N$, in order for the leptogenesis mechanism to work \cite{3311il}, it follows \be m_{\nu_{2,3R}}=-\sqrt{2}h'^\nu_{22,33}\langle \Phi\rangle \sim m_C > m_\Phi\sim m_{\nu_{1R}}=-\sqrt{2}h'^\nu_{11}\langle \Phi\rangle.\ee The inflaton cannot decay into the gauge boson $C$ as well as the heavy right-handed neutrinos $\nu_{2,3R}$ although they have interactions $\mathcal{L}_{\mathrm{int}}\supset 4g^2_N\langle \Phi\rangle \Phi C_\mu C^\mu+(\fr{1}{\sqrt{2}}h'^\nu_{ii}\Phi \bar{\nu}^c_{iR}\nu_{iR}+H.c.)$ for $i=2,3$. However, after the inflation, the inflaton might decay into a pair of scalars ($\chi,S$, and even $\rho,\eta$ if the previous modes are suppressed) or a pair of the light right-handed neutrinos ($\nu_{1R}$) with subsequent thermalization with the standard model particles. The interaction Lagrangian is given by 
\be \mathcal{L}_{\mathrm{int}}\supset \langle \Phi \rangle \Phi \left[\la_{11}\chi^\dagger \chi +\zeta_6 \Tr (S^\dagger S)\right]+\left(\fr{1}{\sqrt{2}} h'^\nu_{11} \Phi \bar{\nu}^c_{1R}\nu_{1R}+H.c.\right).\ee If the inflaton mass is much larger than the products, $m^2_\Phi\gg m^2_{\chi,S,\nu_{1R}}$, the decay rates are 
\be \Ga_\chi= \fr{\la^2_{11}\langle \Phi\rangle^2}{16\pi m_\Phi},\hs \Ga_S= \fr{\zeta^2_{6}\langle \Phi\rangle^2}{16\pi m_\Phi},\hs \Ga_{\nu_{1R}}= \fr{(h'^\nu_{11})^2 m_\Phi }{8\pi}.\ee 

If $\la^2_{11}/\la,\zeta^2_6/\la\gg (h'^\nu_{11})^2$, the inflaton mainly decays into $\chi,S$ with the total width $\Ga=\Ga_\chi+\Ga_S$. The reheating temperature is given by 
\be T_R = \left(\fr{90}{\pi^2g_*}\right)^{1/4}(m_P \Ga)^{1/2}
\simeq 10^{9}\left(\fr{\La}{10^{14}\ \mathrm{GeV}}\right)^{1/2}\left(\fr{10^{-12}}{\la}\right)^{1/4}\fr{(\la^2_{11}+\zeta^2_6)^{1/2}}{10^{-9}}\ \mathrm{GeV},\label{dddch}\ee where $g_*=106.75$ is the effective number of degrees of freedom given at the temperature of the asymmetric production. With the parameters as obtained, taking $(\la^2_{11}+\zeta^2_6)^{1/2}\sim 10^{-9}$ yields $T_R\sim 10^9$ GeV, which is in agreement with the upper bound for the reheating temperature to prevent the gravitino problem \cite{jjeel}. In this case, the right-handed neutrinos may be thermally produced, recognizing a thermal leptogenesis scenario \cite{3311il}.

By contrast, when $\la^2_{11}/\la,\zeta^2_6/\la\ll (h'^\nu_{11})^2$, the inflaton substantially decays into $\nu_{1R}$. The reheating temperature is bounded by 
\be T_R = \left(\fr{90}{\pi^2g_*}\right)^{1/4}(m_P \Ga_{\nu_{1R}})^{1/2}\simeq 1.6 \times 10^{12}\left(\fr{\La}{10^{14}\ \mathrm{GeV}}\right)^{1/2}\left(\fr{\la}{10^{-12}}\right)^{1/4}h'^\nu_{11}\ \mathrm{GeV}.\ee Taking the condition $h'^\nu_{11} \leq \sqrt{\la}\sim 10^{-6}$ and with the other parameters as given, the reheating temperature is limited by $T_R\lesssim 10^6$ GeV, which is significantly smaller than the right-handed neutrino masses. This case realizes a nonthermal leptogenesis scenario since the light right-handed neutrinos are produced by inflaton decay. 

Note that both the cases considered always satisfy the total width to be less than $m_\Phi$ for perturbative decay. On the other hand, the model predicts the value of the reheating temperature to be compatible with thermal productions (see below) after the cosmic inflation \cite{pro3}. Similarly, we
can consider the other single-field inflation scenarios, where the scalar triplet $\chi$ or sextet $S$ plays a role of inflaton.

The above single-field inflation scenarios predict a good approximation to the Gaussian spectrum of primordial fluctuations. The size of non-Gaussian contribution $f_{\mathrm{NL}}$ is suppressed by the slow-roll parameters. However, a combined analysis of the Planck temperature and polarization data shows that $f^{\mathrm{local}}_{\mathrm{NL}}=0.8 \pm 5.0$ \cite{pdg}. There are popular multi-field inflation models which may generate the observably large non-Gaussianity \cite{multi}. It is natural to consider the multi-field inflation in our model due to the presence of a large number of scalar fields behaving in this regime. Let us consider the model where inflation is driven by the multiple scalar consisting of $S_{33}, \chi_3, \phi$ fields. For simplicity, we ignore the soft interaction, $\chi^T S^\dag \chi$, which can be suppressed by some global symmetry. We conveniently define $\phi_1 = \phi^2-\La^2/2$, $\phi_2= S_{33}^2-\Delta^2/2$, $\phi_3 = \chi_3^2 -w^2/2$, and using the potential minimization conditions as obtained in (\ref{addh}). The inflation potential (\ref{infla1}) can be rewritten as follows
\bea
V_{\mathrm{tot}}= \la \phi_1^2+(\zeta_1+\zeta_2) \phi_2^2+\la_2 \phi_3^2+\zeta_6 \phi_1 \ph_2+\la_{11}\phi_1 \phi_3+(\zeta_5+\zeta_8)\phi_2 \phi_3.
\label{mul1}\eea 
The potential given in (\ref{mul1}) contains cross coupling terms between the scalar fields as $\phi_1\phi_2$, $\phi_2\phi_3$, and $\phi_3\phi_1$. In order to make the cross coupling terms disappeared, we can change to the canonical basis system by diagonalizing the corresponding $3\times 3$ matrix. Without loss of generality, we define the new fields as follows
\bea
&& \phi_1^\prime =\frac{-a_{23}\phi_1+\phi_2 +a_{12}a_{23}\phi_3}{1+a_{12}a_{23}a_{31}},\\
&& \phi_2^\prime = \fr{a_{23}a_{31}\phi_1-a_{31}\phi_2+\phi_3}{1+a_{12}a_{23}a_{31}},\\
&& \phi_3^\prime = \frac{\phi_1+a_{12}a_{31}\phi_2-a_{12}\phi_3}{1+a_{12}a_{23}a_{31}}, 
\eea
where
\bea
a_{12}&=&-\frac{ \la (\zeta_5+\zeta_8)^2-\la_{11}^2(\zeta_1+\zeta_2)+\la_2(\zeta_6^2-4\la \zeta_1-4\la \zeta_2)+\sqrt{A}}{2\zeta_6(\zeta_5+\zeta_8)\la-4(\zeta_1+\zeta_2)\la \la_{11}}, \\
a_{23}&=&-\frac{ \la (\zeta_5+\zeta_8)^2+\la_{11}^2(\zeta_1+\zeta_2)-\la_2(\zeta_6^2+4\la \zeta_1+4\la \zeta_2)+\sqrt{A}}{2(\zeta_1+\zeta_2)[(\zeta_5+\zeta_8)\la_{11}-2\zeta_6 \la_2]}, \\
a_{31}&=& \frac{ -\la (\zeta_5+\zeta_8)^2+\la_{11}^2(\zeta_1+\zeta_2)+\la_2(\zeta_6^2-4\zeta_1 \la-4\zeta_2 \la)+\sqrt{A}}{2\la_2(2\la \zeta_5+2\la \zeta_8-\la_{11}\zeta_6)},\\
A &=&-4(\zeta_1+\zeta_2)\la[(\zeta_5+\zeta_8)\la_{11}-2\la_2 \zeta_6]^2 + \left\{(\zeta_1+\zeta_2)\la_{11}^2-\la_2\zeta_6^2\right.\crn
&&\left.+\la[(\zeta_5+\zeta_8)^2-4(\zeta_1+\zeta_2)\la_2]\right\}^2.
\eea
On the basis of the new fields $\phi_1^\prime, \phi_2^\prime, \phi_2^\prime$, the inflation potential in (\ref{mul1}) can be written as
\bea
V_{\mathrm{tot}}&=&[\zeta_1+\zeta_2+a_{31}(\zeta_5+\zeta_8)-a_{31}\la_2]\phi_1^{ \prime 2} +(a_{12}^2 \la +a_{12}\la_{11}+\la_2) \phi_2^{\prime 2}\crn &&+ [a_{23}^2(\zeta_1+\zeta_2)+a_{23}\zeta_6 +\la]\phi_3^{\prime 2}. 
\eea
If $\phi_1^\prime, \phi_2^\prime, \phi_3^\prime$ play a role of inflation, we have a model of multi-field inflation with a separable potential. This inflationary scenario was appropriately considered in \cite{multi-inflation}. The general expression for the nonlinear parameter characterizing non-Gaussianities $f_{NL}$ is suppressed by the number of e-folding for  model with a narrow mass spectrum, and this suppression is enhanced for model with a broad spectrum of masses. We would like to emphasize that in the case, the fields $\phi_1, \phi_2, \phi_3$ play a role of inflation and are non-canonical. In the physical basis, the cross coupling terms between various fundamental scalar fields should be considered, and this might present another source making a large non-Gaussianity.

\section{\label{pheno} Superheavy dark matter}

The recent developments in understanding how matter was created in the early Universe suggests that dark matter might be supermassive. The idea is currently favoured since the well-established, thermal weakly-interacting massive particles have been searched for, but not found. Its mass can be much greater than the weak scale, say the inflation scale, which was created very early, at the end of inflation, in a non-thermal state. It never reached chemical equilibrium with plasma, avoiding the unitarity constraint \cite{unitarity}. A small ratio of thermal energy transferred at the beginning, smaller than $10^{-18}$, suffices to explain the present dark matter abundance, via cosmological mechanisms such as gravitational production \cite{pro9,pro10,pro11,pro12,pro13}, thermal production at reheating \cite{pro1,pro2,pro3,pro4}, non-perturbative parametric resonance effects at preheating \cite{pro5,pro6,pro7}, and topological defects \cite{pro1,pro8}. The last one is irrelevant to this model, while the non-perturbative parametric resonance mechanism is inaccessible since the inflaton couplings to superheavy dark matter always contribute to the Coleman-Weinberg potential which are required to be perturbative as well as retaining the flatness of potential. The remaining mechanisms are viable to be discussed below.       

All the dark matter candidates in this model, say $X^0$, $N_{aR}$, $H'$, $S'_3$, and $A'_3$, have mass proportional to the large scales $\La,w,\Delta$ as the inflation scale. The lightest particle of which (as called LWP) is stabilized by the $W$-parity conservation, responsible for the mentioned superheavy dark matter. When they are created by some source after the end of inflation, they are never to thermalize. The condition for the candidate to lie out of thermal equilibrium and its comoving number density to be constant is that its self-annihilation rate is less than the Hubble parameter, i.e. $n\langle \sigma v \rangle \lesssim H$. Here, the self-annihilation cross-section times the M{\o}ller velocity takes the form $\langle \sigma v \rangle\simeq \al^2_{\mathrm{LWP}}/m^2_{\mathrm{LWP}}$, and $H=\fr{1}{m_P}\sqrt{\fr{V}{3}}\sim \fr{1}{m_P}\sqrt{\la}\La^2$. The condition leads to a bound on the dark matter mass \cite{pro9} \be \fr{m_{\mathrm{LWP}}}{m_P}\gtrsim  10^{-9}\left(\fr{\al_{\mathrm{LWP}}}{\al}\right)^{2/3}\left(\fr{\sqrt{\la}}{10^{-6}}\right)^{1/3}\left(\fr{\La}{m_P}\right)^{2/3}\left(\fr{10^9\ \mathrm{GeV}}{T_R}\right)^{1/3}.\ee Thus, the superheavy dark mater is not to thermalize if its mass satisfies, for instance $m_{\mathrm{LWP}}\gtrsim 10^{9}$~GeV, which is compatible to the inflaton mass as well as those given in the above references. This typical bound intends to change, depending on the self-annihilation coupling $\al_{\mathrm{LWP}}$ and the inflation scenarios to be used. The correct abundance of the candidates is explicitly studied below when we investigate their produced sources.              

The gravitational mechanism is common and model-independent. That being said, due to the interaction of gravitational field with vacuum quantum fluctuations of the dark matter field, our candidate can be generated with an appropriate density, provided that it has a mass proportional to the inflaton mass, $m_{\mathrm{LWP}}\sim m_{\mathrm{inflaton}}\sim 10^{13}$ GeV \cite{pro9,pro10,pro11,pro12,pro13}. In this case, any lightest particle among $X^0$, $N_{aR}$, $H'$, $S'_3$, or $A'_3$ which is identified as LWP is viable. Indeed, considering the single-field inflation scenario with the $B-L$ breaking inflaton field, the present-day density of dark matter is approximated as \cite{shdmde}
\be \Omega_{\mathrm{LWP}}h^2\sim 10^{3}\left(\fr{T_R}{10^{9}\ \mathrm{GeV}}\right)\left(\fr{m_{\mathrm{LWP}}}{10^{13}\ \mathrm{GeV}}\right)^2\left(\fr{m_{\mathrm{LWP}}}{H}\right)^{1/2}e^{-2m_{\mathrm{LWP}}/H}, \ee where $H$ is given at the end of inflation as obtained, $H\sim \fr{\La}{m_P}m_\Phi$. The dark matter mass is proportional to the $B-L$ breaking scale, $m_{\mathrm{LWP}}\simeq g_{\mathrm{LWP}}\La\sim m_\Phi$, thus $m_{\mathrm{LWP}}/H\sim m_P/\La$. The data $\Omega_{\mathrm{LWP}}h^2\sim 0.1$ implies $m_P/\La\gtrsim 6$ for $m_{\mathrm{LWP}}\simeq 3\times 10^{13}$ GeV and $T_R\simeq 10^9$ GeV. Hence, to have the appropriate dark matter density by the gravitational production, the $B-L$ breaking scale $\La$ should be close to the Planck scale.      

This gravitational production mechanism might also affect on the observable quantities such as $r$ and $n_s$ as well as giving rise to considerable isocurvature perturbations. However, the contribution size to the former should be small in comparison to the obtained ones, while the latter would be of interest under the light of the recent experiments, which possibly dedicates a further look to publish elsewhere. For the former, as referred to the single-field inflation scenario with the $B-L$ breaking inflaton field, $\phi$ is a 3-3-1 singlet. It does not interact with $X^0$ gauge boson as well as neutral fermions $N_R$. However, it can interact with the scalar candidates such as $H'$, $S'_3$, and $A'_3$ via the cross coupling constants between scalars in the potential as $\la_{11}$, $\la_{12}$, and $\zeta_6$. As obtained, the interaction strengths $\la_{11},\zeta_6$ are very weak, $\la_{11},\zeta_6\lesssim 10^{-9}$, and $\la_{12}\sim (\la_{11},\zeta_6)$ should be imposed in order to maintain the flatness of the inflaton effective potential since it gets a contribution from the $\la_{12}$ coupling between $\phi,\eta$. Therefore, the effective coupling $a$ is only governed by $g_N$ and $h'^\nu$ as achieved, due to $g^2_N,(h'^\nu)^2\gg \la_{11},\la_{12},\zeta_6$. And, the contributions of the scalar candidates do not affect $r,n_s$ as the effective potential retains unchanged. Additionally, with an appropriate choice of the parameters, the inflaton might decay into the scalar dark matter. But, the contribution to the reheating temperature is at most, only comparable to the one in (\ref{dddch}), which is again in agreement with the existing bounds. Further, the radiation density at the reheating time is given by $\rho_R=(\pi^2/30)g_*T^4_R$, which is not affected too. Correspondingly, the contribution of this density to the number of e-folds, thus to $r,n_s$, is also negligible.                

An alternative interesting origin for LWP results from thermal production during reheating. In this scenario, the radiation is produced as the inflaton decays, but it is only dominated over the universe when the temperature is below the reheating temperature. In fact, the direct decay products of inflaton can rapidly thermalize, forming a plasma with the temperature much beyond the convenient reheating temperature. With this hight temperature background $\sim 10^3 T_R$, the LWP can be created by scattering of light states. Another possibility is that they can be produced directly from the inflaton decay or from the thermalization of the water-fall fields and right-handed neutrinos. 

On the theoretical side, the upper bound for $T_R$ is model-dependent. As referred to App. \ref{apda}, our theory proved is as an alternative to the grand unified theories. In addition, the proton is always stabilized due to $W$-parity as a residual symmetry of the 3-3-1-1 gauge symmetry. There is no reason for the existence of supersymmetry, and thus gravitino. Neglecting this obstacle, the reheating temperature may be raised much higher. For this case, the LWP can be produced by thermal fusions, e.g. from radiations too.                 

To find the LWP relic density, it is necessarily to solve the system of Boltzmann equations describing the redshift and interchange in the energy densities for components, including the inflaton density, the radiation density, and the LWP density. Generalizing the result in \cite{pro3}, the present LWP density is given as
\bea
       \Omega_{\mathrm{LWP}} h^2 = m_{\mathrm{LWP}}^2 \langle \sigma v\rangle
        \left(\frac{g_\ast}{200} \right)^{\frac{-3}{2}}\left(
        \frac{2000 T_{\mathrm{R}}}{m_{\mathrm{LWP}}}
        \right)^7,
   \eea
where $g_\ast$ is the effective number of degrees of freedom in the radiation, and $\langle \sigma v\rangle$ is the thermal averaged LWP annihilation cross-section times the M{\o}ller velocity. We would like to stress again that all the LWPs are heavy with masses proportional to $\La, \Delta, w$ as the inflation scale, and that these candidates are electrically neutral and colorless as expected. Thus, this ensures that the LWP is stable and is an suitable candidate for superheavy dark matter, thermally produced in the $10^3T_R$ scale. 

Depending on the parameter space, we have the following possibilities:
\ben
\item The LWP is the $W$-odd gauge boson, $X^0$. The appearance of the scalar sextet does not contribute to
the mass of the neutral, complex gauge boson as well as its couplings to the standard model particles.
Hence, the annihilation of $X^0$ into the standard model
particles is analogous to those in \cite{3311dm}, where the dominant contribution is the channel, $X^0X^{0*}\rightarrow W^+ W^-$. The
thermal average of the annihilation cross-section times the velocity was evaluated as \cite{3311dm}
\bea
\langle \sigma v\rangle_{X} \simeq \frac{5\al^2 m_X^2}{8 s_W^4 m_W^4}.
\eea
Hereafter, we take $\frac{\al^2}{(150^2\ \mathrm{GeV}^2)} \simeq 1$ pb and $s_W^2 \simeq 0.23$. Additionally, the reheating temperature as calculated in the previous section is $T_{R} \lesssim 10^9$ GeV, so we choose $T_R=10^9$ GeV. Hence, the present density of $X$ gauge boson is
 \bea
\Omega_X h^2 \simeq 6.0340 g_*^{-\frac{3}{2}} \left(\frac{10^{27}\ \mathrm{GeV}}{m_X} \right)^3.
 \eea It is evident that $g_* \sim 100$--200 and $m_X$ is limited below the Planck scale. Hence, $X$ cannot be a candidate for dark matter since it overpopulates, $\Omega_X h^2\gg 1$.
\item The LWP is the lightest neutral fermion among $N_{aR}$, denoted as $N_R$. The fermions $N_R$ annihilate into the standard model particles due to the contribution of the new neutral gauge bosons $Z'$ and $Z''$ via s-channels as well as the new complex gauge bosons $X,Y$ via t-channels. Here, the annihilation modes into $Z,H,t,\tau,\nu_\tau$ where the leptons have t-channels are dominated. Using the limit $m_{N_R}\gg m_t, m_Z, m_{H}\gg m_{lep}$, the
thermal average of the annihilation cross-section times the velocity is approximated as \cite{3311dm}
\bea
\langle \sigma v\rangle_{N_R}  \simeq \frac{\al^2}{(150\ \mathrm{GeV})^2}
\frac{(2557.5\ \mathrm{GeV})^2m^2_{N_R}}{m^4_{Z^\prime}},
\eea with the assumption that $m_X\sim m_Y\sim \fr{\sqrt{3-t^2_W}}{2}m_{Z'}\sim \fr{\sqrt{3-t^2_W}}{2}m_{Z''}$.  
In this case, the relic density of the fermion dark matter can be written as
\bea
   \Omega_{N_{R}} h^2 \simeq \frac{6.08121}{g_*^{3/2}z^4}
  \left(  \frac{10^{13}\ \mathrm{GeV}}{m_{N_{R}}} \right)^7,
\eea
where $z\equiv \frac{m_{Z^\prime}}{m_{N_{R}}}\sim 1$ since these masses are both proportional to the large scale, $\La,\Delta, w$. The correct density as observed demands $m_{N_R}\sim 10^{13}$ GeV.
\item The LWP is a scalar among the $H',S'_3,A'_3$ states, by which we choose $H'$ for investigation. The annihilation cross-section of $H'$ into the standard model particles
was obtained in \cite{3311ph} by the Higgs portal as follows 
 \bea
 \langle \sigma v \rangle_{H'} \simeq \frac{\al^2}{(150\ \mathrm{GeV})^2} \la^\prime \left(
 \frac{1.328\ \mathrm{GeV}}{m_{H^\prime}}
 \right)^2,
 \eea
 where $\la^\prime$ is the effective coupling between two $H^\prime$ scalars with
 two standard model Higgs bosons. The $H^\prime$ relic density is
 \bea
\Omega_{H^\prime} h^2 \simeq 1.63967 \la^\prime
 g_*^{-\frac{3}{2}} \left(\frac{10^{12}\ \mathrm{GeV}}{m_{H^\prime}} \right)^7.
\eea
Given that the scalar coupling is proportional to one, $\la'\sim1$, the observed abundance of dark matter is recovered if $m_{H'}\sim 10^{12}$ GeV.
 \een

\section{\label{concl}Conclusions}

We have shown that the 3-3-1-1 model can work under the three distinct regimes of the energy scale, characterized by the VEVs as $\kappa\sim m_\nu $, $(u,v)\sim m_{W,Z}$, and $(w,\Delta,\La)\sim m_{\mathrm{inflaton}}$. The $B-L$ breaking scale, $\La$, is responsible for the type I seesaw mechanism and inflation scenario, so it is naturally picked up a value in the large energy regime. The introduction of the scalar sextet implies that the 3-3-1 breaking scales, $\Delta$ and $w$, are also large, proportional to $\La$, recognizing the fact that the $B-L$ and 3-3-1 symmetries are nontrivially unified. Therefore, the new physics regime of the 3-3-1-1 model is actually realized in the inflation scale as governed by $(w,\Delta,\La)$. The consistent smallness of $\kappa$ and neutrino masses are ensured by the type I and II seesaw mechanisms as a result of the 3-3-1-1 gauge symmetry breaking, and that they are naturally suppressed by the large scales. In other words, the conventional seesaw mechanisms can be manifestly explained by a noncommutative $B-L$ dynamics associated with the 3-3-1 gauge symmetry. And, the resulting 3-3-1-1 model provides not only the neutrino masses and leptogenesis but also the other consequences behind such as inflation scenarios and superheavy dark matter.

The 3-3-1-1 breaking fields can behave as inflatons deriving the inflationary expansion of the early universe. The several single-field inflation scenarios have been interpreted, in which the case associated with the $B-L$ breaking was explicitly shown, taking the contribution of the superheavy particles to the inflaton effective potential. The inflaton can have a mass in the $10^{8}$--$10^{12}$ GeV order corresponding to $\La=10^{14}$--$10^{18}$ GeV. The reheating temperature is naturally bounded by $10^{9}$ GeV if the inflaton decays into the scalars or by a lower value if it decays into the right-handed neutrinos. The multi-field inflation scenarios can be explicitly implemented in this model as cooperated by the superheavy Higgs fields, but their contribution to the isocurvature and non-Gaussian perturbations should be small due to the slow-role approximation. When turning on the coupling terms between inflatons, these effects may be enhanced, which was not evaluated by this work.               

The breakdown of the 3-3-1-1 gauge symmetry induces the $W$-parity, i.e. $R$-parity, as a residual gauge symmetry, making the $W$-particles that carry abnormal $B-L$ number to be odd. The $W$-particles include a non-Hermitian gauge boson $X^0$, scalars $H'$, $S'_3$, $A'_3$, and fermions $N_{aR}$ besides the other electrically-charged states, which all have mass proportional to the large scales ($\La, w,\Delta$). The lightest $W$-particle or the LWP is stabilized, responsible for superheavy dark matter. The $X^0$ as LWP can only be gravitationally produced, with a mass $m_X\sim 10^{13}$ GeV. Alternatively, the LWP as a lightest neutral fermion among $N_{1,2,3R}$ or a neutral scalar among $H'$, $S'_3$, and $A'_3$ can be created in the early universe by either the gravitational or thermal productions, which depend on their mass in the $10^{13}$--$10^{12}$ GeV range or possibly lower according to the thermal mechanism. The contribution of superheavy dark matter to the slow-roll parameters $r,n_s$ and the reheating temperature is negligible. However, their effects for the isocurvature and non-Gaussian perturbations may be considerable in comparing to the mentioned multi-field inflation scenarios.     

Conclusively, the 3-3-1-1 model at the large energy regime recognizes an actual unification of the $B-L$ and 3-3-1 symmetries, yielding the potential solution to the important issues of particle physics and cosmology, such as neutrino masses, baryon asymmetry, dark matter, and inflation. Although it is presented as an alternative to the grand unified theories, at an extremely high energy regime, a possible unification of the gauge couplings along with a more-fundamental gauge symmetry might emerge, to be devoted for further studies \cite{stringcom}.

\section*{Acknowledgments}

This research is funded by Vietnam National Foundation for Science and Technology Development
(NAFOSTED) under grant number 103.01-2016.77.

\appendix

\section{\label{ano3311} Anomaly checking}

With the $X,N$ charges as given in the text, we have 
\bea [SU(3)_C]^2U(1)_X &\sim& \sum_{\mathrm{quarks}} (X_{q_L}-X_{q_R}) = 3X_{Q_3}+2\times 3 X_{Q_\al}-3X_{u_a}-3X_{d_a}-X_{U}-2X_{D_\al}\crn
&=&3\left(1/3\right)+6\left(0\right)-3\left(2/3\right)-3\left(-1/3\right)-\left(2/3\right)-2\left(-1/3\right)=0. \eea
\bea [SU(3)_C]^2U(1)_N &\sim&  \sum_{\mathrm{quarks}} (N_{q_L}-N_{q_R}) = 3N_{Q_3}+2\times 3 N_{Q_\al}-3N_{u_a}-3N_{d_a}-N_{U}-2N_{D_\al}\crn
&=& 3\left(2/3\right)+6\left(0\right)-3\left(1/3\right)-3\left(1/3\right)-\left(4/3\right)-2\left(-2/3\right)=0. \eea
\bea 
[SU(3)_L]^2 U(1)_X &\sim& \sum_{\mathrm{(anti)triplets}} X_{F_L}= 3X_{\psi_a}+3X_{Q_3}+2\times 3 X_{Q_\al}\crn
 &=& 3\left(-1/3\right)+3\left(1/3\right)+6\left(0\right)=0.\eea  
\bea 
[SU(3)_L]^2 U(1)_N &\sim& \sum_{\mathrm{(anti)triplets}} N_{F_L}= 3N_{\psi_a}+3N_{Q_3}+2\times 3 N_{Q_\al} \crn
&=& 3\left(-2/3\right)+3\left(2/3\right)+6\left(0\right)=0.  \eea 
The last two anomalies have taken the color number, i.e. $3$'s in the second and last terms, into account, and below the presence of this number should be understood. Note also that for antitriplets, we have $\Tr[(-T^*_i)(-T^*_j)X]=\Tr[T_i T_j X]$ and similarity for $N$ charge.    
\bea [\mathrm{gravity}]^2U(1)_X&\sim&\sum_{\mathrm{fermions}}(X_{f_L}-X_{f_R})=3\times 3 X_{\psi_a}+3\times 3 X_{Q_3}+2\times 3 \times 3 X_{Q_\al}\crn
&&-3\times 3 X_{u_a}-3\times 3 X_{d_a}-3X_{U}-2\times 3 X_{D_\al}-3X_{e_a}-3X_{\nu_a}\crn
&=&3\times 3 (-1/3)+3\times 3 (1/3)+2\times 3\times 3 (0)-3\times 3 (2/3)\crn
&&-3\times 3 (-1/3) -3(2/3)-2\times 3 (-1/3)-3(-1)-3(0)=0.\eea
\bea [\mathrm{gravity}]^2U(1)_N&\sim&\sum_{\mathrm{fermions}}(N_{f_L}-N_{f_R})=3\times 3 N_{\psi_a}+3\times 3 N_{Q_3}+2\times 3 \times 3 N_{Q_\al}\crn
&&-3\times 3 N_{u_a}-3\times 3 N_{d_a}-3N_{U}-2\times 3 N_{D_\al}-3N_{e_a}-3N_{\nu_a}\crn
&=&3\times 3 (-2/3)+3\times 3 (2/3)+2\times 3\times 3 (0)-3\times 3 (1/3)\crn
&&-3\times 3 (1/3) -3(4/3)-2\times 3 (-2/3)-3(-1)-3(-1)=0.\eea
\bea [U(1)_X]^2U(1)_N&=&\sum_{\mathrm{fermions}}(X^2_{f_L}N_{f_L}-X^2_{f_R}N_{f_R})=3\times 3 X^2_{\psi_a}N_{\psi_a}+3\times 3 X^2_{Q_3} N_{Q_3}\crn &&+2\times 3\times 3 X^2_{Q_{\al}}N_{Q_{\al}}-3\times 3 X^2_{u_a}N_{u_a}-3\times 3 X^2_{d_a}N_{d_a}-3X^2_{U} N_{U}\crn
&&-2\times 3 X^2_{D_\al}N_{D_\al}-3X^2_{e_a} N_{e_a}-3X^2_{\nu_a}N_{\nu_a}\crn
&=&3\times 3 (-1/3)^2(-2/3)+3\times 3 (1/3)^2(2/3)+2\times 3\times 3 (0)^2(0)\crn
&&-3\times3(2/3)^2(1/3)-3\times 3(-1/3)^2(1/3)-3(2/3)^2(4/3)\crn
&&-2\times3(-1/3)^2(-2/3)-3(-1)^2(-1)-3(0)^2(-1)=0. \eea 
\bea U(1)_X[U(1)_N]^2&=&\sum_{\mathrm{fermions}}(X_{f_L}N^2_{f_L}-X_{f_R}N^2_{f_R})=3\times 3 X_{\psi_a}N^2_{\psi_a}+3\times 3 X_{Q_3} N^2_{Q_3}\crn &&+2\times 3\times 3 X_{Q_{\al}}N^2_{Q_{\al}}-3\times 3 X_{u_a}N^2_{u_a}-3\times 3 X_{d_a}N^2_{d_a}-3X_{U} N^2_{U}\crn
&&-2\times 3 X_{D_\al}N^2_{D_\al}-3X_{e_a} N^2_{e_a}-3X_{\nu_a}N^2_{\nu_a}\crn
&=&3\times 3 (-1/3)(-2/3)^2+3\times 3 (1/3)(2/3)^2+2\times 3\times 3 (0)(0)^2\crn
&&-3\times3(2/3)(1/3)^2-3\times 3(-1/3)(1/3)^2-3(2/3)(4/3)^2\crn
&&-2\times3(-1/3)(-2/3)^2-3(-1)(-1)^2-3(0)(-1)^2=0. \eea 
\bea [U(1)_X]^3&=&\sum_{\mathrm{fermions}}(X^3_{f_L}-X^3_{f_R})=3\times 3 X^3_{\psi_a}+3\times 3 X^3_{Q_3}+2\times 3\times 3 X^3_{Q_\al}\crn
&&-3\times 3 X^3_{u_a}-3\times 3 X^3_{d_a}-3X^3_{U}-2\times 3 X^3_{D_\al}-3X^3_{e_a}-3X^3_{\nu_a}\crn
&=&3\times 3 (-1/3)^3+3\times 3 (1/3)^3+2\times 3\times 3 (0)^3-3\times3(2/3)^3\crn
&&-3\times 3(-1/3)^3-3(2/3)^3-2\times3(-1/3)^3-3(-1)^3-3(0)^3=0.\eea  
\bea [U(1)_N]^3&=&\sum_{\mathrm{fermions}}(N^3_{f_L}-N^3_{f_R})=3\times 3 N^3_{\psi_a}+3\times 3 N^3_{Q_3}+2\times 3\times 3 N^3_{Q_\al}\crn
&&-3\times 3 N^3_{u_a}-3\times 3 N^3_{d_a}-3N^3_{U}-2\times 3 N^3_{D_\al}-3N^3_{e_a}-3N^3_{\nu_a}\crn
&=&3\times 3 (-2/3)^3+3\times 3 (2/3)^3+2\times 3\times 3 (0)^3-3\times3(1/3)^3\crn
&&-3\times 3(1/3)^3-3(4/3)^3-2\times3(-2/3)^3-3(-1)^3-3(-1)^3=0.\eea

\section{\label{apda}GUT embedding}

Let  us study the possibility of embedding the 3-3-1-1 model into a grand unified theory. We assume that the 3-3-1-1 model can be unified by a simple group such as $SU(n)$ or $SO(n)$. Of course the gauge group $SU(3)_C \otimes SU(3)_L \otimes U(1)_X \otimes U(1)_N$ should be a subgroup of the grand unified group, and thus the fermion
content in our model is included in the matter representations of the
grand unified group. Since the 3-3-1-1 group is embedded into the simple group,
the $X$ and $N$ charges are determined as a combination
of the Cartan generators of the grand unified group. Therefore, the $X$ and $N$
generators are traceless. It means that the total $X$ and
$N$ charges in every matter representation of the grand unified group
must add up to zero. Furthermore, if we look at the representation of each
fermion multiplet given in Eqs. (\ref{fe1}--\ref{fermion}), we see that the $X$
and $N$ charges have different values, by which we cannot arrange
the fermion multiplets in the 3-3-1-1 model into 
the matter representations of the unified group so that the total $X$ and
$N$ charges simultaneously add up to zero. 

To embed our fermion content into the representations of the unified group, we have to rearrange 
the fermion representations of the original model as in \cite{grand}, or otherwise we introduce new fermion
multiplets. Let us illustrate this by selecting a grand unified
group $SU(7) \supset$ 3-3-1-1 group. The anomaly-free combination of $SU(7)$ representations is $7^* + 21+ 35^*$. Each irreducible representation of $SU(7)$ can decompose into 
irreducible representations of 3-3-1-1 group, which depends on the choice of the intermediate subgroup. Let us consider two cases. The first case is that $SU(7)\rightarrow SU(6) \otimes U(1)_a  \rightarrow SU(3)\otimes SU(3) \otimes U(1)_b \otimes U(1)_a$. The representations, $7^*+21+35^*$, decompose into the 3-3-1-1 representations, 
\bea 7^*+21+35^*&=& \left [(1,1,0,6a) \oplus (1,3^*, 3b, -a) \oplus (3^*, 1, -3b, -a) \right] + \left[ (3^*,1,6b,2a) 
\right.\crn && \oplus \left. (3,3,0,2a) \oplus (1,3^*, -6b, 2a) \oplus (3,1,3c,-5a) \oplus (1,3,-3c,-5a) \right] \crn &&+
\left [ (1,3,6b,4a) \oplus (3^*, 3^*, 0, 4a) \oplus (3,1,-6b,4a)  \oplus (1,1,9c,-3a) \right.\crn && \oplus \left. (1,1,-9c,-3a)
\oplus (3^*,3,3d,-3a) \oplus (3,3^*,-3d,-3a)\right].
\label{aa}\eea
The second case is if $SU(7)\rightarrow SU(4) \otimes SU(3) \otimes U(1)_a  \rightarrow SU(3)\otimes SU(3) \otimes U(1)_b \times U(1)_a$, the representations, $7^*+21+35^*$, decompose into 3-3-1-1 subgroups as follows
\bea
7^*+21+35^*& = & \left[ (1,3^*,0,4a) \oplus (1,1,3c,-3a) \oplus (3^*,1,-c,-3a)\right] + \left[(3^*,1,2b,6a) \right.\crn && \oplus \left. (3,1,-2b,6a) \oplus (3,3,c,-a) \oplus (1,3,-3c,-a) \oplus (1,3^*,0,-8a) \right] \crn &&+ \left[(1,1,0,12a) \oplus (1,3,3c,5a)
\oplus(3^*,3,-c,5a) \oplus(3,3^*,2d,-2a) \right.\crn && \oplus \left. (3,3^*,-2d,-2a) \oplus(3,1,e,-9a) \oplus(1,1,-3e,-9a) \right].
\label{bb}\eea
Due to a separation of $SU(7)$ representations given in (\ref{aa}) or (\ref{bb}), it is easily to see that there is no choice the value of $(a,b,c,d,e)$ so that each $SU(7)$ representation contains at least  two different fermion multiplets of the 3-3-1-1 model. Therefore, if embedding the 3-3-1-1 model into $SU(7)$, the number of irreducible representations of $SU(7)$ must equal to that of the present fermion multiplets containing in the 3-3-1-1 model. It means that for each fermion family, we have to introduce three sets of $7^*+21+35^*$ representations of the $SU(7)$ group. So the fermion content in the $SU(7)$ grand unified model will appear much more new 
fermions beyond the 3-3-1-1 model.  This is not favourite choice, and other issue may arise because the QCD asymptotic freedom
is not ensured due to a largely-numerical contribution of new quark fields. 

To avoid appearance of much more new fermions in the SU(7) unified theory, we have to change the quantum numbers $X$ and $N$ for each fermion multiplets or in other words we have to modify the fermion content in the 3-3-1-1 model as similarly done in the 3-3-1 models \cite{grand}. There may be other option that the  interactions and their gauge symmetries need not necessarily be unified at a grand unified scale. The 3-3-1-1 symmetry as it stands is enough to describe the physics at such large scale.

\end{document}